\documentclass[oneside]{article}
\usepackage[T1]{fontenc}
\usepackage{authblk}
\usepackage{float}
\usepackage{amsmath}
\usepackage{graphicx}
\usepackage{xfrac}
\usepackage[english]{babel}
\usepackage{lmodern}
\usepackage[T1]{fontenc}
\usepackage[latin9]{inputenc}
\PassOptionsToPackage{normalem}{ulem}
\usepackage{ulem}
\usepackage{mathtools}
\usepackage{graphicx}
\usepackage{esint}
\usepackage[unicode=true,
 bookmarks=false,
breaklinks=false,pdfborder={0 0 1},backref=false,colorlinks=false]
 {hyperref}
\usepackage[a4paper, total={210mm,297mm},margin=2.0cm]{geometry}
\usepackage{breakurl}
\usepackage[normalem]{ulem}
\usepackage{color}

\usepackage{datetime}
\newdate{date}{13}{04}{2015}

\providecommand{\tabularnewline}{\\}

\title{JETSPIN: a specific-purpose open-source software for simulations of nanofiber electrospinning}

\author[1]{Marco Lauricella \thanks{Electronic address: \texttt{m.lauricella@iac.cnr.it}; Corresponding author}}
\author[1]{Giuseppe Pontrelli}
\author[2]{Ivan Coluzza}
\author[3,4]{Dario Pisignano}
\author[1]{Sauro Succi}

\affil[1]{Istituto per le Applicazioni del Calcolo CNR, Via dei Taurini 19, 00185 Rome, Italy}
\affil[2]{Faculty of Physics, University of Vienna, Boltzmanngasse 5, 1090 Vienna, Austria}
\affil[3]{Dipartimento di Matematica e Fisica Ennio De Giorgi,University of Salento, via Arnesano, 73100 Lecce, Italy}
\affil[4]{Istituto Nanoscienze-CNR, Euromediterranean Center for Nanomaterial Modelling and Technology (ECMT), via Arnesano, 73100 Lecce, Italy}

\date{\displaydate{date}}

\begin{document}

\maketitle
 
\begin{abstract}
We present the open-source computer program JETSPIN, specifically
designed to simulate the electrospinning process of nanofibers. 
Its capabilities are shown with proper reference to the underlying
model, as well as a description of the relevant input variables 
and associated test-case simulations. 
The various interactions included in the electrospinning model implemented in JETSPIN
are discussed in detail. The code is designed to exploit different
computational architectures, from single to parallel processor workstations.
This paper provides an overview of JETSPIN, focusing primarily on its
structure, parallel implementations, functionality, performance, and availability.
\end{abstract}

\section{Introduction}

In the recent years, electrospun nanofibers have gained a considerable 
interest due to many potential  industrial applications, such as tissue engineering,
air and water filtration, drug delivery and regenerative medicine.
In particular, the high surface area to material mass ratio of the fibers offers 
intriguing prospects for technological applications. 
As consequence, several studies have focused on the characterization and production
of uni-dimensionally elongated nanostructures. 
A number of reviews \cite{li2004electrospinning,greiner2007electrospinning,carroll2008nanofibers, huang2003review, persano2013industrial}
and books \cite{yarin2014fundamentals,wendorff2012electrospinning,pisignanoelectrospinning}
concerning electrospinning have been published in the last decade. 

Typically, electrospun nanofibers are produced at laboratory scale via 
the uniaxial stretching of a jet, which is ejected at a nozzle from an electrified polymer
solution. Indeed, the initial elongation of the jet can be obtained 
by applying an external electrostatic field
between the spinneret and a conductive collector. Electrospinning
involves mainly two sequential stages in the uniaxial elongation
of the extruded polymer jet: an initial quasi-steady stage, in which
the electric field stretches the jet in a straight path away from
the nozzle, and a second stage characterized
by a bending instability induced by small perturbations, which misalign
the jet away from its axis of elongation \cite{zeng2006numerical}. 
These small disturbances may originate from mechanical vibrations 
at the nozzle or  from hydrodynamic-aerodynamic
perturbations within the experimental apparatus. Such a misalignment provides
an electrostatic-driven bending instability before the jet reaches
the conductive collector, where the fibers are finally collected.
As a consequence, the jet path length between the nozzle and the collector
increases and the stream cross-section undergoes a further decrease.
The ultimate goal of the electrospinning process are to control the cross-sectional radius
and to maximize the uniformity of the collected fibers. 
By a simple argument of mass conservation, this is tantamount to 
maximizing the jet length by the time it reaches
the collecting plane. By the same argument, it is therefore of interest
to minimize the length of the initial stable jet region. 
Consequently, the bending instability is a desirable effect, as it produces a higher
surface-area-to-volume ratio of the jet, which is transferred to the
resulting nanofibers \cite{feng2003stretching}. 

Recently, due to the broad interest of nanotechnology and to the wide application fields
gained by polymer nanofibers even at industrial scale \cite{persano2013industrial},
electrospinning has attracted the attention of a large community of researchers, including
modeling and computational aspects \cite{reneker2000bending,yarin2001taylor,fridrikh2003controlling,theron2004experimental,lu2006computer,zeng2006numerical}.
In fact, computational models provide a useful tool to elucidate the physics
of electrospinning and provide information which may be used 
for the design of new electrospinning experiments. 
Numerical simulations can also help improving
the capability of predicting the role of the key process parameters
and exert a better control on the resulting nanofiber structure.
Although some authors used dissipative particle dynamics mesoscale simulation methods \cite{wang2013simulation}, various models treat the jet filament 
as a series of discrete elements ({\em beads}) obeying the equations of continuum
mechanics \cite{reneker2000bending,yarin2001taylor}. 
Each bead is subject to different types
of interactions, such as long-range Coulomb repulsion, viscoelastic drag,
the force related to the external electric field, and so on. 
The main aim of such models is to capture the complexity of the resulting 
dynamics and to provide the set of parameters driving the process.
The effect of fast-oscillating loads on the bending instability has been also explored in an extensive computational study \cite{coluzza2014ultrathin}.

In recent works, we extended the unidimensional bead-spring model,
developed by Pontrelli \textit{et al.} \cite{pontrelli2014electrospinning},
to include a nonlinear dissipative-perturbing force which models the
effects of the air drag force. This has been accomplished by adding  a random
and a dissipative force to the equations of motion.
In particular, we investigated both linear and non-linear Langevin-like models 
to describe air drag effects \cite{lauricella2014electrospinning,lauricella2015langevin}.

Encompassing the previous efforts, in this paper we present, 
along with the overall model, a detailed algorithm and the corresponding FORTRAN code, 
JETSPIN, specifically designed to simulate 
the electrospinning process under a variety of different conditions and experimental 
settings. This comprehensive platform is devised in such a way to handle 
a variety of different cases via a suitable choice of the input variables.
The framework is developed to exploit several computational architectures, both 
serial and parallel. With most of parameters taken from 
relevant literature in the field, a number of test cases have been carried out 
and an excellent agreement with experiments has been found. \par

JETSPIN, as open-source software, can be used to carry out a systematic sensitivity analysis 
over a broad range of parameter values. The results of simulations provide valuable 
insight on the physics of the process and can be used to assess
experimental procedures for an optimal design of the equipment and to control 
processing strategies for technologically advanced nanofibers.

\section{Structure \label{sec:Structure}}

JETSPIN is written in free format FORTRAN90, and it consists of approximately
140 subroutines. The source exploits the modular approach provided
by the programming language. All the variables having in common description
of certain features or method are grouped in modules. The convention
of explicit type declaration is adopted, and all the arguments passed
in calling sequences of functions or subroutines have defined intent.
We use the PRIVATE and PUBLIC accessibility attributes in order to
decrease error-proneness in programming. 

\begin{figure}[H]
\begin{centering}
\includegraphics[scale=0.1]{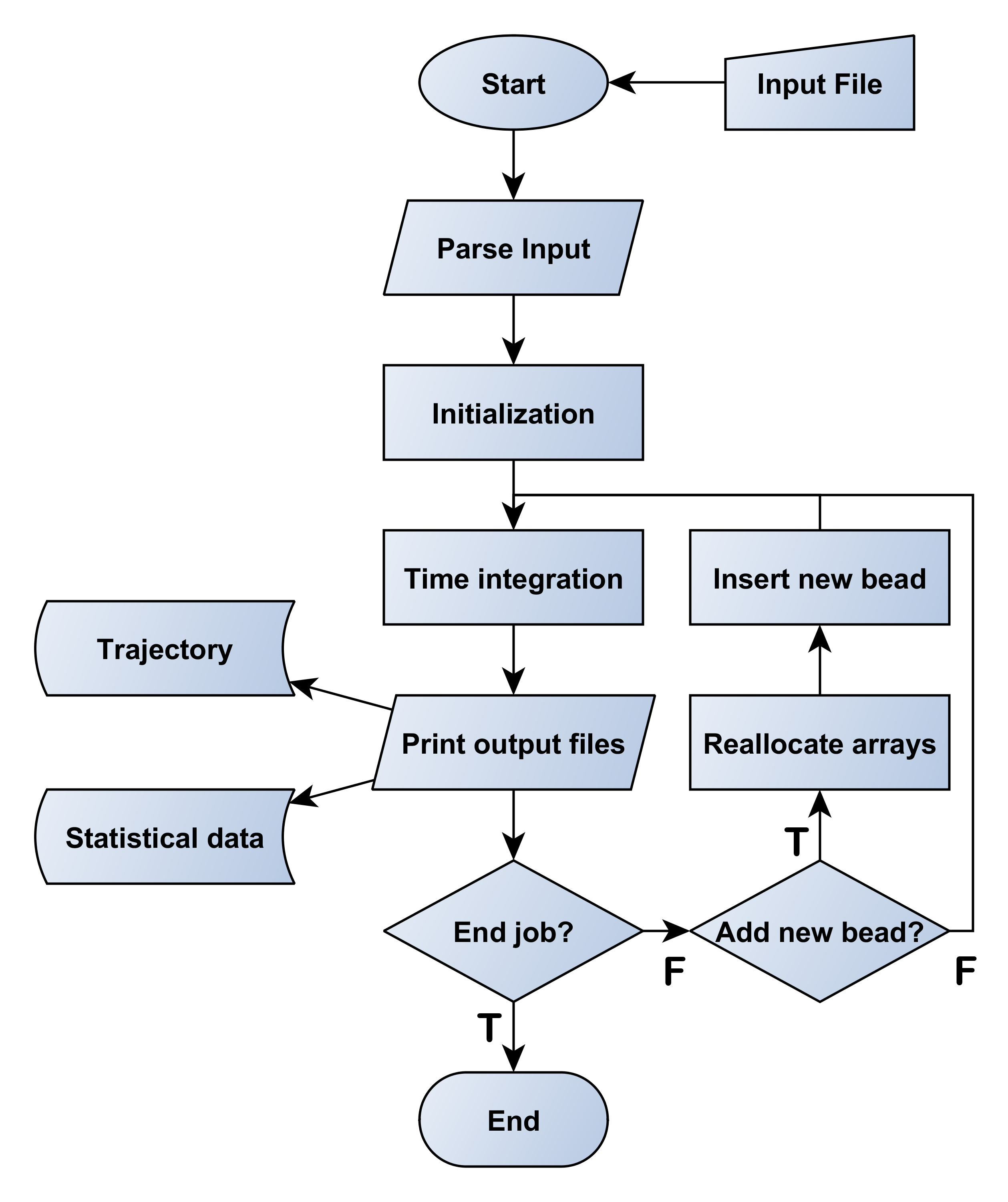}
\par\end{centering}

\protect\caption{Structure of the main JETSPIN program.}

\label{fig:scheme-jetspin}
\end{figure}

The main routines have been gathered in the \texttt{main.f90} file,
which drives all the CPU-intensive computations needed for the capabilities
mentioned below. The variables describing the main features of jets
(position, velocity, etc.) are declared in the \texttt{nanojet\_mod.f90}
file, which also contains the main subroutines for the memory management
of the fundamental data of the simulated system. Since the size system
is strictly time-dependent, JETSPIN exploits the dynamic array allocation
features of FORTRAN90 to assign the necessary array dimensions. In
particular, the size system is modified by the routines \texttt{add\_jetbead}
and \texttt{erase\_jetbead}, while the decision of the main array
size, declared as \texttt{mxnpjet}, is handled by the routine \texttt{reallocate\_jet}.
The sizes of various service bookkeeping arrays are handled within
a parallel implementation strategy, which exploits few dedicated subroutines
(see Sec \ref{sec:Parallelization}). All the implemented time integrators
are written in the \texttt{integrator\_mod.f90} file, which contains
the routine \texttt{driver\_integrator} to select the proper integrator,
as indicated in the input file. All the terms of equations of motion
for the implemented model (see Sec \ref{sec:Overview-of-model}) are
computed by routines located in the \texttt{eom\_mod.f90} file, which
call other subroutines in the files \texttt{coulomb\_force\_mod.f90},
\texttt{viscoelastic\_force\_mod.f90} and \texttt{support\_functions\_mod.f90}.
A summarizing scheme of the main JETSPIN program in the \texttt{main.f90}
file has been sketched in Fig\ref{fig:scheme-jetspin}.

The user can carry out simulations of electrospun jets without a detailed
understanding of the structure of JETSPIN code. All the parameters
governing the system can be defined in the input file (see Sec \ref{sec:Description-of-the-input}),
which is read by routines located in the \texttt{io\_mod.f90} and
\texttt{parse\_mod.f90} files. Instead, the user should be acquainted
with the model described in Sec \ref{sec:Overview-of-model}. The
content of the output file is completely customizable by the input
file as described in Sec \ref{sec:Description-of-the-input}, and
it can report different time-averaged observables computed by routines
of the module \texttt{statistic\_mod} (see Sec \ref{sec:Description-of-the-output}).
The routines in the \texttt{error\_mod.f90} file can display various
warning or error banners on computer terminal, so that the user can
easily correct the most common mistakes in the input file.

JETSPIN is supplied as a single UNIX compressed (tarred and gzipped)
directory with four sub-directories. All the source code files are
contained in the \textit{source} sub-directory. The \textit{examples}
sub-directory contains different test cases that can help the user
to edit new input files. The \textit{build} sub-directory stores a
UNIX \texttt{makefile} that assembles the executable versions of
the code both in serial and parallel version with different compilers.
Note that JETSPIN may be compiled on any UNIX platform. The \texttt{makefile}
should be copied (and eventually modified) into the \textit{source} sub-directory, where
the code is compiled and linked. A list of targets for several common workstations
and parallel computers can be used by the command "\texttt{make target }", where \texttt{target} is one of the options reported in Tab \ref{Tab:targets}. 
On Windows system we advice the user to compile JETSPIN under the command-line interface Cygwin \cite{racine2000cygwin}.
Finally, the binary executable file
can be run in the \textit{execute} sub-directory.

\section{Parallelization\label{sec:Parallelization}}

The parallel infrastructure of JETSPIN incorporates the necessary
data distribution and communication structures. The parallel strategy
underlying JETSPIN is the Replicated Data (RD) scheme\cite{smith1991molecular},
where fundamental data of the simulated system are reproduced on all
processing nodes. In simulations of electrospinning, the fundamental data
consist of position, velocity, and viscoelastic stress arrays at each
bead in which the jet is discretised (see below Sec \ref{sec:Overview-of-model}).
Further data defining mass and charge of each bead are also replicated.
However, all auxiliary data are distributed in equal portion of data
(as much as possible) for each processor. Despite 
other parallel strategies being available such as the Domain Decomposition\cite{brown1993domain},
our experience has shown that such volume of data is by no means prohibitive
on current parallel computers.

By the RD scheme, we implement the following parallel procedure:
1) A set of arrays (in the following text referred to as global arrays) containing position, velocity, viscoelastic stress, mass and charge of each bead  is replicated on each processing node. 
2) The routine \texttt{set\_chunk} distributes the computational work over all the nodes 
by assigning a nanofiber chunk to each node. In particular, the first and last beads of the jet chunk dealt by the $i-th$ node are declared as \texttt{mystart} and \texttt{myend},
which have different values for each node.  
3) Each node evolves in time the nanofiber for its assigned chunk of jet. Thus, the set of global arrays are updated only for the jet chunk which is handled by the specific node, 
while all the remaining values are set to zero. 
At this stage each node exploits service bookkeeping arrays which are not replicated, and whose size is dynamically allocated
on the basis of the chunk size in order to save memory space.
4) Finally, global summation routines are employed in order to make the updated data of the global arrays available to all nodes.
Note that we adopt in JETSPIN a simple strategy
of communication between nodes, which is handled by global summation
routines.

The module \texttt{version\_mod} located in the \texttt{parallel\_version\_mod.f90}
file contains all the global communication routines which exploit
the MPI (Message Passing Interface) library. It is worth stressing that a FORTRAN90
compiler and an MPI implementation for the specific machine architecture
are required in order to compile JETSPIN in parallel mode. An alternative
version of the module \texttt{version\_mod} is located in the \texttt{serial\_version\_mod.f90}
file, and it can be easily selected by appropriate targets in the \texttt{makefile}
at compile time (see Tab \ref{Tab:targets}). By selecting this version, JETSPIN can also be run
on serial computers without modification, even though the code has been designed
to run on parallel computers. 

The size system is strictly time-dependent as mentioned in Sec \ref{sec:Structure}
and, therefore, the memory of various service bookkeeping arrays is
dynamically distributed over all the processing nodes. In particular,
the bookkeeping array size, declared as \texttt{mxchunk}, is managed
by the routine \texttt{set\_mxchunk}. All the beads of the discretised
jet are assigned at every time step to a specific node by the
routine \texttt{set\_chunk}, and their temporary data are stored
in bookkeeping arrays belonging to the assigned node. It is worth
stressing that the communication latency makes the parallelization
efficiency strictly dependent on the system size. Therefore, we only
advice the use of JETSPIN in parallel mode whenever the user expects a system
size with at least 50 beads for each node (further details in Sec
\ref{sec:Performance}).

\section{Overview of model\label{sec:Overview-of-model}}

\subsection{Equations of motion}

The model implemented in JETSPIN is an extension of the Lagrangian
discrete model introduced by Reneker et al. \cite{reneker2000bending}. The model provides
a compromise of efficiency and accuracy by representing the filament
as a series of $n$ beads (jet beads) at mutual distance $l$ connected
by viscoelastic elements. The length $l$ is typically larger than
the cross-sectional radius of the filament, but smaller than the characteristic
lengths of other observables of interest (e.g. curvature radius).
Each $i-th$ bead has mass $m_{i}$ and charge $q_{i}$ 
(not necessarily equal for all the beads). 
Evaporation has been neglected since
it is not expected to introduce qualitative changes the jet dynamics \cite{reneker2000bending}.
However, the effect of solvent evaporation likely leads to a slight solidification of the jet, 
altering the rheological parameters of the polymer solution. This latter issue was addressed by 
an ad-hoc evaporation model proposed by Yarin et al. \cite{yarin2001bending}, 
whose implementation in JETSPIN
will be considered in future releases.
The jet is modelled
as a viscoelastic Maxwell fluid, so that the stress $\sigma_{i}$
for the element connecting bead $i$ with bead $i+1$
is given by the viscoelastic constitutive equation:

\begin{equation}
\frac{d\sigma_{i}}{dt}=\frac{G}{l_{i}}\frac{dl_{i}}{dt}-\frac{G}{\mu}\sigma_{i},\label{eq:stress-ode}
\end{equation}
where $l_{i}$ is the length between the bead $i$
with the bead $i+1$, $G$ is the elastic modulus, $\mu$ is the viscosity
of the fluid jet, and $t$ is time. Given $a_{i}$ the cross-sectional
radius of the filament at the bead $i$, the viscoleastic force $\vec{\textbf{f}}_{\upsilon e}$
pulling bead $i$ back to $i-1$ and towards $i+1$ is

\begin{equation}
\vec{\textbf{f}}_{\upsilon e,i}=-\pi a_{i}^{2}\sigma_{i}\cdot\vec{\textbf{t}}_{i}+\pi a_{i+1}^{2}\sigma_{i+1}\cdot\vec{\textbf{t}}_{i+1}\text{,}
\end{equation}
where $\vec{\textbf{t}}_{i}$ is the unit vector pointing bead $i$
from bead $i-1$. The surface tension force $\vec{\textbf{f}}_{st}$
for the $i-th$ bead is given by

\begin{equation}
\vec{\textbf{f}}_{st,i}=k\cdot\pi\left(\frac{a_{i}+a_{i-1}}{2}\right)^{2}\alpha\cdot\vec{\textbf{c}}_{i}\text{,}
\end{equation}
where $\alpha$ is the surface tension coefficient, $k$ is the local
curvature, and $\vec{\textbf{c}}_{i}$ is the unit vector pointing
the center of the local curvature from bead $i$. Note the force $\vec{\textbf{f}}_{st}$
is acting to restore the rectilinear shape of the bending part of
the jet. 

In the electrospinning experimental configuration an intense electric potential $V_{0}$ is applied between the spinneret and a conducting
collector located at distance $h$ from the injection point. As consequence,
each $i-th$ bead undergoes the electric force 

\begin{equation}
\vec{\textbf{f}}_{el,i}=e_{i}\frac{V_{0}}{h}\cdot\vec{\textbf{x}}\text{,}
\end{equation}
where $\vec{\textbf{x}}$ is the unit vector pointing the collector
from the spinneret assuming a vertical $x$ axis starting at the spinneret ($x=0$). Note that in Ren
model the intense electric potential $V_{0}$  is assumed to be static in order to
avoid the computationally expensive integration of Poisson equation,
whereas in reality $V_{0}$ is depending on the net charge of the jet
so as to maintain constant the potential at the electrodes. The latter
issue was elegantly addressed by Kowalewski et al. \cite{kowalewski2009modeling},
and its implementation in JETSPIN will be planned.
Furthermore, a model using a lattice method for electromagnetic wave propagation
\cite{hanasoge2011lattice,benzi1992lattice,ottaviani1990numerical} is planned to be implemented in future releases in order to
deal with electrospinning process in the presence of oscillating electric fields.

The net Coulomb force $\vec{\textbf{f}}_{c}$ acting on the $i-th$
bead from all the other beads is given by

\begin{equation}
\vec{\textbf{f}}_{c,i}=\sum_{\substack{j=1\\
j\neq i
}
}^{n}\frac{q_{i}q_{j}}{R_{ij}^{2}}\cdot\vec{\textbf{u}}_{ij}\text{,}
\end{equation}
where $R_{ij}=\left[\left(x_{i}-x_{j}\right)^{2}+\left(y_{i}-y_{j}\right)^{2}+\left(z_{i}-z_{j}\right)^{2}\right]^{1/2}$,
and $\vec{\textbf{u}}_{ij}$ is the unit vector pointing the $i-th$ bead
from $j-th$ bead.. Although the Reneker model provides a reasonable description
for the spiral motion of the jet, the last term $\vec{\textbf{f}}_{c}$
introduces mathematical inconsistencies due to the discretization
of the fiber into point-charges. Indeed, the charge induces a field
on the outer shell of the fiber and not on the center line (as in
the implemented model). Different approaches were developed to overcome
this issue which usually imply strong approximations\cite{feng2002stretching,feng2003stretching,hohman2001electrospinning}.
Other strategies use a less crude approximation by accounting for
the actual electrostatic form factors between two interacting sections
of a charged fiber\cite{kowalewski2005experiments} or involving more
sophisticated methods which exploit the tree-code hierarchical force
calculation algorithm\cite{kowalewski2009modeling}. The implementation
of methods based on tree-code hierarchical force calculation
algorithm in JETSPIN will be considered in future.

Although usually much smaller than the other driving forces, the body
force due to the gravity is computed in the model by the usual expression

\begin{equation}
\vec{\textbf{f}}_{g,i}=m_{i}g\cdot\vec{\textbf{x}},
\end{equation}
where $g$ is the gravitational acceleration.

The combined action of these forces governs the elongation of the
jet according to the Newton's equation:

\begin{equation}
m_{i}\frac{d\vec{\boldsymbol{\upsilon}}_{i}}{dt}=\vec{\textbf{f}}_{el,i}+\vec{\textbf{f}}_{c,i}+\vec{\textbf{f}}_{\upsilon e,i}+\vec{\textbf{f}}_{st,i}+\vec{\textbf{f}}_{g,i}\:\text{,}\label{eq:force-EOM}
\end{equation}
where $\vec{\boldsymbol{\upsilon}}_{i}$ is the velocity of the $i-th$
bead. The velocity $\vec{\boldsymbol{\upsilon}}_{i}$ satisfies the
kinematic relation:

\begin{equation}
\frac{d\vec{\textbf{r}}_{i}}{dt}=\vec{\boldsymbol{\upsilon}}_{i}\label{eq:pos-EOM}
\end{equation}
where $\vec{\textbf{r}}_{i}$ is the position vector of the $i-th$
bead, $\vec{\textbf{r}}_{i}=x_{i}\vec{\textbf{x}}+y_{i}\vec{\textbf{y}}+z_{i}\vec{\textbf{z}}$.
The three Eqs \ref{eq:stress-ode}, \ref{eq:force-EOM} and \ref{eq:pos-EOM}
form the set of equations of motion (EOM) governing the time evolution
of system.

Depite the experimental evidence that the air drag affects the jet dynamics \cite{spinning1991science}, 
the effects of aerodynamics are neglected at this stage.
An extended stochastic model, recently developed in Refs \cite{lauricella2014electrospinning,lauricella2015langevin},
including air drag effects is already planned for the next version
of JETSPIN.

\subsection{Perturbations at the nozzle}

The spinneret nozzle is represented by a single mass-less point of
charge $\bar{q}$ fixed at $x=0$ (nozzle bead). Its charge $\bar{q}$
is assumed equal to the mean charge value of the jet beads. Such charged
point can be also interpreted as a small portion of jet which is fixed
at the nozzle. In JETSPIN it is possible to add small perturbations
to the $y_{n}$ and $z_{n}$ coordinates of the nozzle bead in order
to model fast mechanical oscillations of the spinneret\cite{coluzza2014ultrathin}.
Given the initial position of the nozzle 

\begin{subequations}

\begin{equation}
y_{n}=A\cdot\cos\left(\varphi\right)
\end{equation}

\begin{equation}
z_{n}=A\cdot\sin\left(\varphi\right),
\end{equation}

\end{subequations}

the equations of motion for the nozzle bead are

\begin{subequations}

\begin{equation}
\frac{dy_{n}}{dt}=-\omega\cdot z_{n}
\end{equation}

\begin{equation}
\frac{dz_{n}}{dt}=\omega\cdot y_{n},
\end{equation}

\end{subequations}

where $A$ denotes the amplitude of the perturbation, while $\omega$
and $\varphi$ are its frequency and initial phase, respectively.

\subsection{Jet insertion\label{sub:Jet-insertion}}

The jet insertion at the nozzle is modelled as follows. For sake
of simplicity, let us consider a simulation which starts with only
two bodies: a single mass-less point fixed at $x=0$ representing the
spinneret nozzle, and a bead modelling an initial jet segment of mass
$m_{i}$ and charge $e_{i}$ located at distance $l_{step}$ from
the nozzle along the $x$ axis. Here, $l_{step}$ denotes the length
step used to discretise the jet in a sequence of beads. The starting
jet bead is assumed to have a cross-sectional radius $a_{0}$, defined
as the radius of the filament at the nozzle before the stretching
process. Applying the condition of conservation of the jet volume,
the relation $\pi a_{i}^{2}\, l_{i}=\pi a_{0}^{2}l_{step}$ is valid
for any $i-th$ bead. Furthermore, the starting jet bead has an initial
velocity $\upsilon_{s}$ along the $x$ axis equal to the bulk fluid
velocity in the syringe needle. Once this traveling jet bead is a
distance of $2\cdot l_{step}$ away from the nozzle, a new jet bead
(third body) is placed at distance $l_{step}$ from the nozzle along
the straight line joining the two previous bodies. Let us now label
$i-1$ the farthest bead from the nozzle, and $i$ the last inserted
bead. The $i-th$ bead is inserted with the initial velocity $\upsilon_{i}=\upsilon_{s}+\upsilon_{d}$,
where $\upsilon_{d}$ denotes the dragging velocity computed as

\begin{equation}
\upsilon_{d}=\frac{\upsilon_{i-1}-\upsilon_{s}}{2}.
\end{equation}

Here, the dragging velocity should be interpreted as an extra term
which accounts for the drag effect of the electrospun jet
on the last inserted segment. Note that the actual dragging velocity
definition was chosen in order to not alter the strain velocity term
$\left(1/l_{i-1}\right ) \cdot \left(dl_{i-1}/dt\right )$ of Eq \ref{eq:stress-ode} before
and after the bead insertion.

\subsection{Dimensionless quantities\label{sub:Introduction-of-dimensionless}}

In JETSPIN all the variables are automatically rescaled and stored
in dimensionless units. In order to adopt a dimensionless form of
the equations of motion, we use the dimensionless scaling procdure
proposed by Reneker et al.\cite{reneker2000bending}. We define a characteristic
length

\begin{equation}
L_{0}=\sqrt{\frac{\bar{q}^{2}}{\pi a_{0}^{2}G}}=l_{step}\sqrt{\frac{\pi a_{0}^{2}\rho_{V}^{2}}{G}},\label{eq:Length-scale}
\end{equation}
where we write the charge $q$ as $\pi a_{0}^{2}l_{step}\rho_{V}$,
denoting $\rho_{V}$ the electric volume charge density of the filament.
Further, we divide the time $t$ and the stress $\sigma$ by their
respective characteristic scales reported in Tab\ref{tab:tabella-adimensionale}.
By using the volume conservation condition, and introducing the dimensionless
variables in EOM, we obtain:

\begin{subequations}

\begin{equation}
\frac{d\vec{\bar{\textbf{r}}}_{i}}{d\bar{t}}=\vec{\boldsymbol{\bar{\upsilon}}}_{\, i}\label{eq:EOM-A1}
\end{equation}

\begin{equation}
\frac{d\bar{\sigma}_{i}}{d\bar{t}}=\frac{1}{\bar{l}_{i}}\frac{d\bar{l}_{i}}{d\bar{t}}-\bar{\sigma}_{i}\label{eq:EOM-A2}
\end{equation}

\begin{equation}
\begin{alignedat}{1}\frac{d\vec{\boldsymbol{\bar{\upsilon}}}_{\, i}}{d\bar{t}}= & V\cdot\vec{\textbf{x}}+\sum_{\substack{j=1\\
j\neq i
}
}^{n}\frac{Q_{ij}}{\bar{R}_{ij}^{2}}\cdot\vec{\textbf{u}}_{ij}-L_{step}F_{ve,i}\frac{\bar{\sigma}_{i}}{\bar{l}_{i}}\cdot\vec{\textbf{t}}_{i}+L_{step}F_{ve,i+1}\frac{\bar{\sigma}_{i+1}}{\bar{l}_{i+1}}\cdot\vec{\textbf{t}}_{i+1}+\\
 & +L_{step}A_{i}\frac{\bar{k}}{4}\left(\frac{1}{\sqrt{\bar{l}_{i}}}+\frac{1}{\sqrt{\bar{l}_{i-1}}}\right)^{2}\cdot\vec{\textbf{c}}_{i}+F_{g}\cdot\vec{\textbf{x}}
\end{alignedat}
\label{eq:EOM-A3}
\end{equation}

\end{subequations}

where we used the dimensionless derived variables and groups defined
in Tab\ref{tab:tabella-adimensionale}. It is worth stressing that
the viscoelastic and surface tension force terms are slightly different
from the dimensionless form provided by Reneker et al. \cite{reneker2000bending},
since we are considering the most general case $l_{step}\neq L.$ 

Similarly, the equations of motion of the nozzle become

\begin{subequations}

\begin{equation}
\frac{d\bar{y}_{i}}{d\bar{t}}=-K_{s}\cdot\bar{z}_{i}
\end{equation}

\begin{equation}
\frac{d\bar{z}_{i}}{d\bar{t}}=K_{s}\cdot\bar{y}_{i}\:,
\end{equation}

\end{subequations}

with the dimensionless parameter $K_{s}$ defined in Tab\ref{tab:tabella-adimensionale}.

\subsection{Integration schemes}

In order to integrate the homogeneous differential EOM
we discretise time as $t_{i}=t_{0}+i\Delta t$ with $i=1,\ldots,n_{steps}$,
where $n_{steps}$ denotes the number of sub-intervals. In JETSPIN
three different integration schemes can be exploited: the first-order accurate Euler
scheme, the second-order accurate Heun scheme (sometimes called second-order
accurate Runge-Kutta), and the fourth-order accurate Runge-Kutta scheme\cite{press2007numerical}.
The user can select a specific scheme by using appropriate keys in the
input file, as described in Sec \ref{sec:Description-of-the-input}.
In addition, the time step $\Delta t$ is automatically rescaled
by the quantity $\tau$, in accordance with the mentioned dimensionless
scaling convention.

\section{Description of input file \label{sec:Description-of-the-input}}

In order to run JETSPIN simulations an input file has to be prepared,
which is free-form  with no sequence field and case-insensitive. The input file has to
be named \texttt{input.dat}, and it contains the selection of the
model system, integration scheme directives, specification of various
parameters for the model, and output directives. The input file does
not requires a specific order of key directives, and it is read by
the input parsing module. Every line is treated as a command sentence
(record). Records beginning with the symbol \# (commented) and blank
lines are not processed, and may be added to aid readability. Each
record is read in words (directives and additional keywords and numbers),
which are recognized as such by separation by one or more space characters. 

As in the example given in \texttt{input test}, the last record is
a \texttt{finish} directive, which marks the end of the input data.
Before the \texttt{finish} directive, a wide list of directives may
be inserted (see Appendix A). The key \texttt{systype} should be
used to set the jet model. In JETSPIN two models are available:
1) the one dimensional model similar to the model of Sec \ref{sec:Overview-of-model}
but assuming the jet to be straight along the $\vec{\textbf{x}}$
axis, and, therefore, neglecting the surface tension force $\vec{\textbf{f}}_{st}$;
2) the three dimensional model described in Sec \ref{sec:Overview-of-model}.
Internally these options are handled by the integer variable \texttt{systype},
which assumes the values explained in Appendix A. Further details
of the 1-D model can be found in Refs \cite{pontrelli2014electrospinning,carroll2011discretized}.
A series of variables are mandatory and have to be defined. For example,
\texttt{timestep}, \texttt{final time}, \texttt{initial length},
etc. (see underlined directives in Appendix A). A missed definition
of any mandatory variable will call an error banner on the terminal.
Given the mandatory directive \texttt{initial length}, the user 
can define the initial jet geometry in two ways:
1) the discretization step length is specified by the
directive \texttt{resolution}, which causes automatically the setting
of the jet segments number (the number of segment in which the jet
is discretised); 2) the jet segments number is declared by the directive
\texttt{points}, while the value of the discretization step length
is automatically set by the program (as in Example \textbf{input 3}).
The directive \texttt{cutoff} indicates the length of the upper and
lower proximal jet sections, which interact via Coulomb force on
any bead. It is worth stressing that the length value is set at the
nozzle, so that the effective cutoff increases along the simulation
as the jet is stretched.

The user should pay special attention in choosing the time integration
step given by the directive \texttt{timestep}, whose detailed
considerations are provided in Subsec \ref{sub:Numerical accuracy}.
The reader is referred to Appendix A for a complete listing of all
directives defining the electrospinning parameters. Not all
these quantities are mandatory, but the user is informed that whenever
a quantity is missed, it is usually assumed equal to zero by default
(exceptions are stressed in Appendix A). Note that in the current software version all the quantities have
to be expressed in centimeter\textendash gram\textendash second unit
system (e.g. charge in statcoulomb, electric potential in statvolt,
etc.).

\section{Description of output files \label{sec:Description-of-the-output}}

A series of specific directives causes the writing of output files
(see Appendix A). In JETSPIN two output files can be written: 1) the
file \texttt{statdat.dat} containing time-dependent statistical data
of simulated process; 2) the file \texttt{traj.xyz} reporting the
jet trajectory in XYZ file format. 

Various statistical data can be written on the file \texttt{statdat.dat}
, and the user can select them in input using the directive \texttt{printstat list}
followed by appropriate symbolic strings, whose list with corresponding
meanings is reported in Appendix B. The statistical observables will
be printed as mean values averaged over the time interval indicated by the directive \texttt{print time}, 
and reported on the same line following the order specified
in input file. In the same way, a list of statistical data can be
printed on computer terminal using the directive \texttt{print list}
followed by the symbolic strings of Appendix B.

The file \texttt{traj.xyz} is written as a continuous series of XYZ
format frames taken at time interval, so that it can be read by suitable visualization 
programs (e.g. VMD-Visual Molecular Dynamics \cite{humphrey1996vmd},
UCSF Chimera \cite{pettersen2004ucsf}, etc.) to generate animations.
The number of elements contained in the file is kept constant equal
to the value specified by the directive \texttt{print xyz maxnum},
since few programs (e.g. VMD) do not manage a variable number of elements
along simulations. If the actual number of beads is lower than
the given constant \texttt{maxnum}, the extra elements are printed
in the origin point.

\section{Numerical tests}

Here, we show three different examples of simulations. Each example
addresses a specific issue of the jet model: 1) the choice of
a suitable time step for the integration scheme; 2) the choice of
a suitable length step for the jet discretization in order to properly
approach the continuum jet description; 3) Fidelity of the
model in reproducing experimental data.

\subsection{Numerical accuracy and time step\label{sub:Numerical accuracy}}

\begin{figure}[H]
\begin{centering}
\includegraphics[scale=0.5]{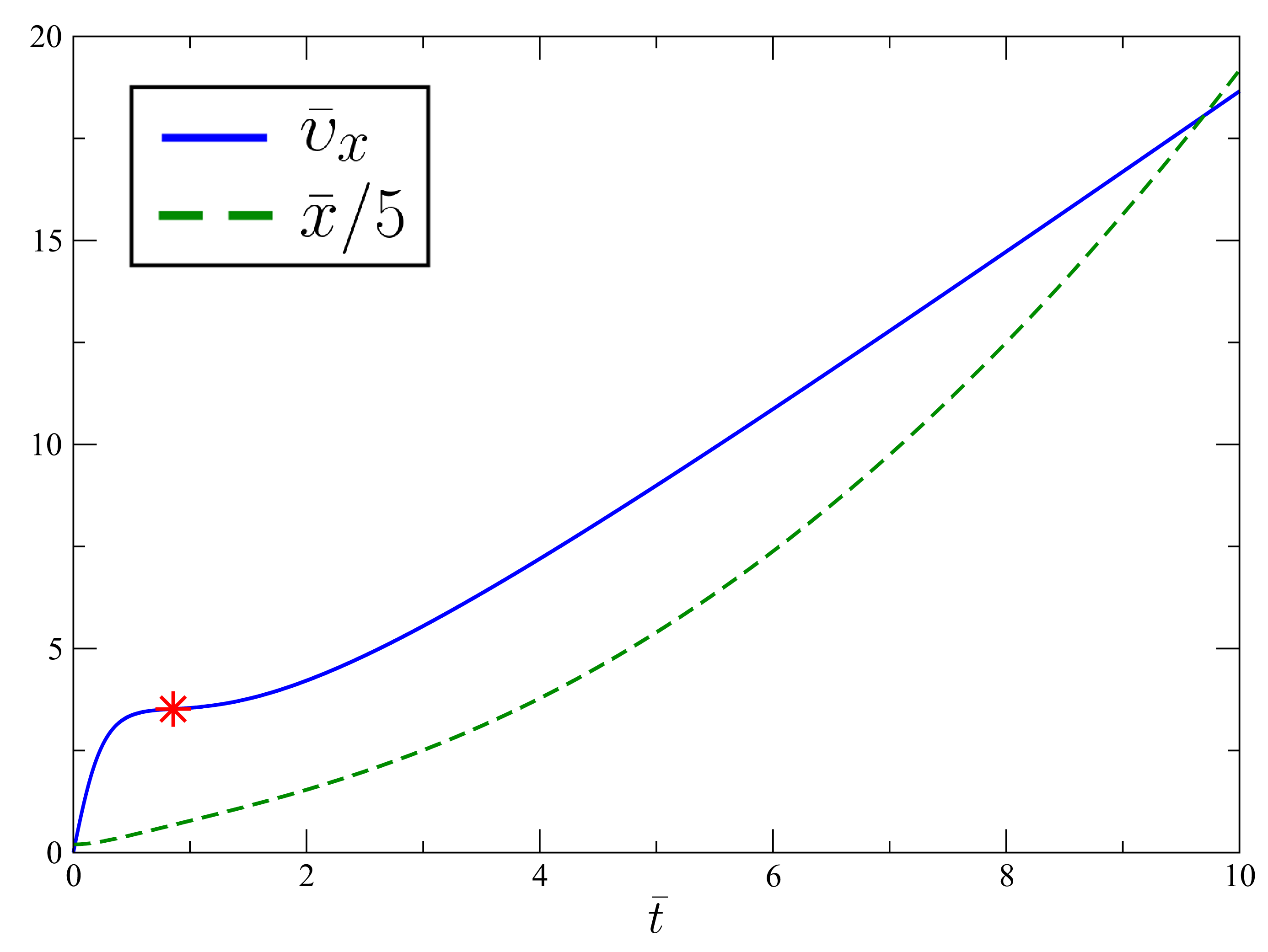}
\par\end{centering}

\protect\caption{Time evolution of the velocity $\bar{\upsilon}_{x}$ (continuous line)
and the length $\bar{x}$ rescaled by a factor $1/5$ (dotted line).
Two stages of the elongation process are observed. The first stage
comes to a quasi stationary point (denoted by a red star symbol).
Then, in the second stage, the velocity comes to a near linearly increasing
regime. The characteristic time and length scales are equal
to $0.01\,\text{s}$ and $0.319\,\text{cm}$, respectively.}

\label{Fig:time-evol-7.1}
\end{figure}

\begin{figure}[H]
\begin{centering}
\includegraphics[scale=0.5]{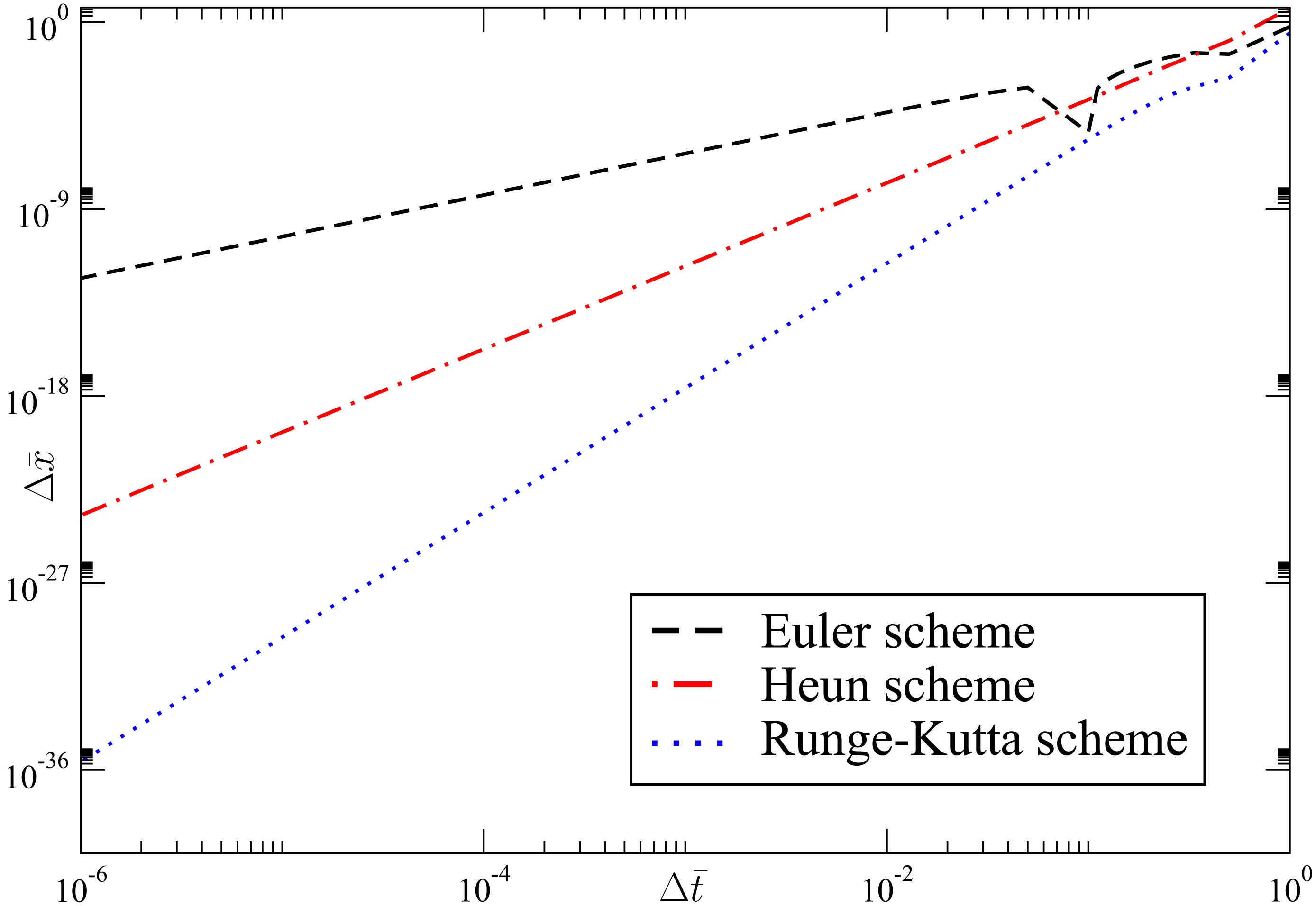}
\par\end{centering}

\protect\caption{Deviation of the conserved quantity $\Delta\bar{x}$ versus time step
$\Delta\bar{t}$ in log-log plot obtained by using the Euler scheme
(black line), Heun scheme (red line), and Runge-Kutta scheme (blue
line). The characteristic time and length scales are equal
to $0.01\,\text{s}$ and $0.319\, \text{cm}$, respectively.}

\label{Fig:accuracy-7.1}
\end{figure}

Now, we intend to assess a typical time step value, $\Delta t$, for
the integration schemes implemented in JETSPIN. To this purpose, exploiting
the time reversibility and using dimensionless quantities, we integrate
Eqs \ref{eq:EOM-A1}, \ref{eq:EOM-A2} and \ref{eq:EOM-A3} forward
for $n_{steps}$ time steps, and backward for further $n_{steps}$
time steps in the interval $\bar{t}_{a}=0$ and $\bar{t}_{b}=5$.
Finally, we compute the average absolute error

\begin{equation}
\Delta\bar{x}=\left(\frac{1}{2 n_{steps}}\right) \left|\bar{x}_{2n_{steps}}-\bar{x}_{0}\right|,\label{eq:cons-obs}
\end{equation}
where $\bar{x}$ is the dimensionless position (defined in Subsec
\ref{sub:Introduction-of-dimensionless}) along the $x$ axis of the
bead (describing the jet) from the nozzle (located at zero). Here,
$\bar{x}_{0}$ and $\bar{x}_{2n_{steps}}$ denote respectively the
position of the jet bead at the beginning and at the end of the time
integration. Note a perfect integrator ideally recovers $\bar{x}_{0}$
after $2n_{steps}$ time steps, so that we should obtain $\Delta\bar{x}=0$.
The procedure was performed on the input example labeled \textbf{input
1}, and it was repeated for different values of time step $\Delta\bar{t}$.
This input file provides the dimensionless parameter values $Q=12$,
$V=2$ and $F_{ve}=12$, which have been already used as reference
case in Refs \cite{reneker2000bending,pontrelli2014electrospinning}.
All the simulations start with the initial conditions $\bar{x}=1$,
$\bar{\sigma}=0$, and $\bar{\upsilon}_{x}=0$. For sake of completeness,
we report in Fig\ref{Fig:time-evol-7.1} the time evolution of the
position $\bar{x}$ and velocity $\bar{\upsilon}_{x}$. As already
noted in Refs \cite{reneker2000bending,pontrelli2014electrospinning},
we identify two sequential stages in the elongation process. In the
first regime, we observe a little increase of $\bar{\upsilon}_{x}$
which rises up to achieve a quasi stationary point, where the viscoelastic
force balances the sum of the Coulomb and electric forces (due to
the external electric field), providing a nearly zero value of the
total force. Then, in the second stage the velocity trend comes to
a near linearly increasing regime. In Fig\ref{Fig:accuracy-7.1}
we report the logarithmic trend of $\Delta\bar{x}$ versus $\Delta\bar{t}$
for the three different integration schemes. Here, the characteristic
time and length scales are equal to $0.01\,\text{s}$ and $0.319\,\text{cm}$,
respectively. As expected, we note a

 precision of the Runge-Kutta
scheme, while the computational cost increases by increasing the accuracy
of the integration scheme, as discussed in more detail in Sec \ref{sec:Performance}.
The Euler's method shows a lower numerical accuracy at larger values
of $\Delta\bar{t}$, in particular close to $\Delta\bar{t}=10^{-1}$.
The Heun scheme provides a compromise
of efficiency and accuracy, keeping the absolute error $\Delta\bar{x}$
lower than $10^{-12}$ already for time step $\Delta\bar{t}=10^{-2}$.
However, a good practice would be to perform preliminary tests of accuracy
and efficiency for any specific case, since the accuracy is dependent
on the magnitude of the dimensionless parameters.

\subsection{Discretization length step}

The discretization length step plays an important role, since we describe
a continuous material system (the jet) by a series of discrete bodies.
In particular, decreasing the discretization length step $l_{step}$,
we approach the continuous description of the problem, and, therefore,
an asymptotic behavior should be observed. 

\begin{figure}[H]
\begin{centering}
\includegraphics[scale=0.5]{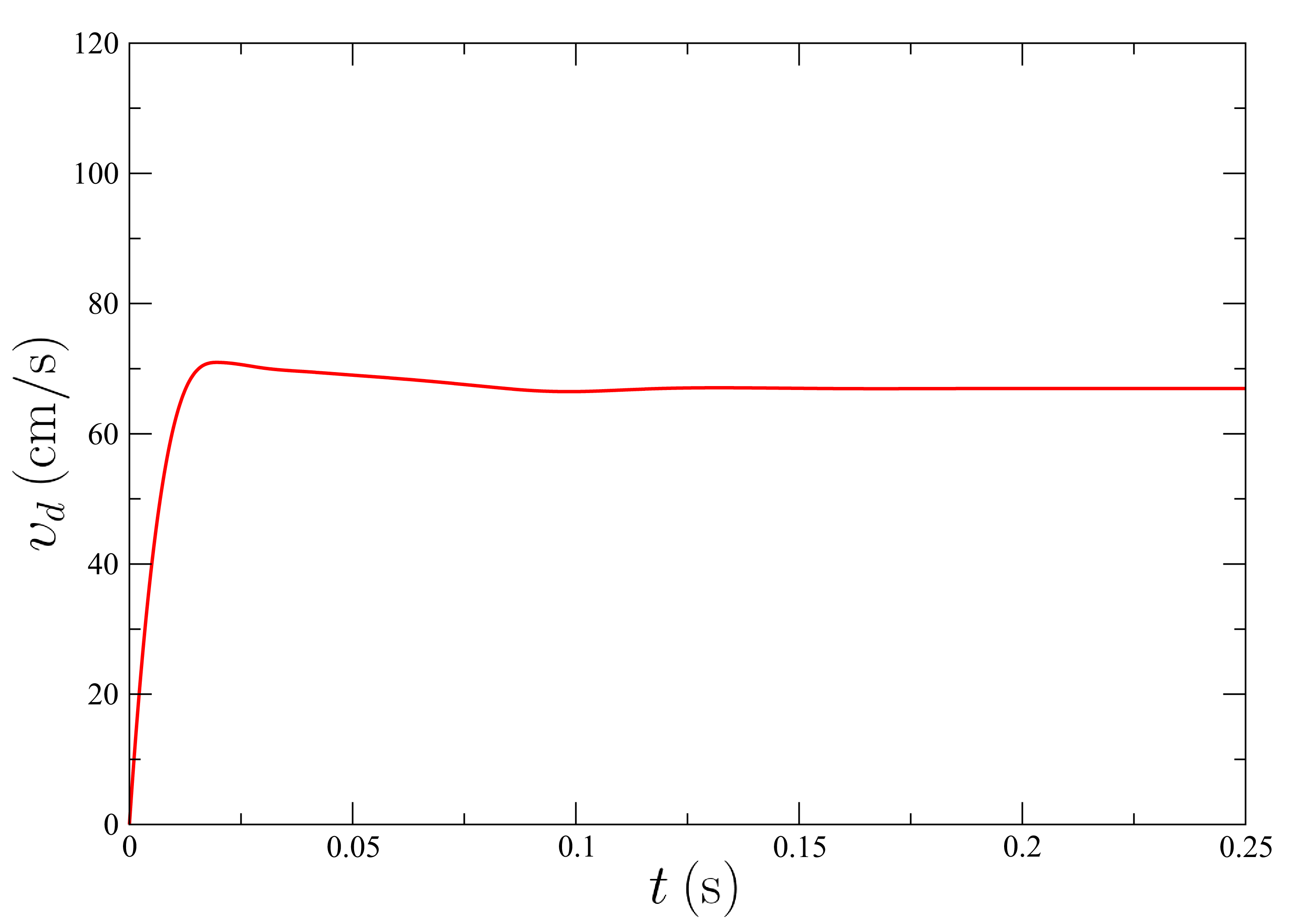}
\par\end{centering}

\protect\caption{Time evolution of the drag velocity $\upsilon_{d}$. After
an initial drift the $\upsilon_{d}$ value reaches a stationary regime
with minimal fluctuations around a mean value.}

\label{Fig:dragvel-7.2}
\end{figure}

\begin{figure}[H]
\begin{centering}
\includegraphics[scale=0.12]{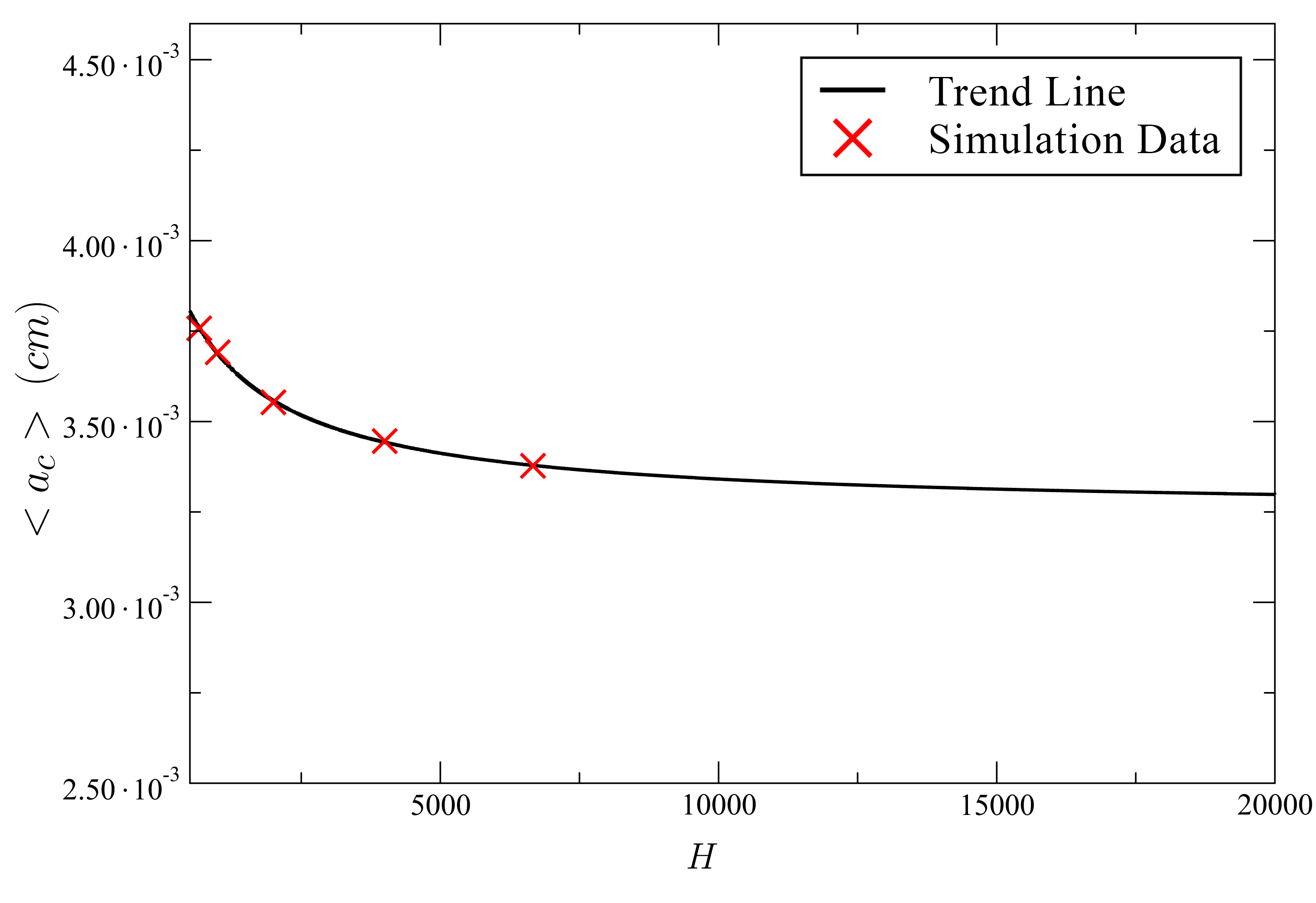}
\par\end{centering}

\protect\caption{Averaged cross section radius $<a_{c}>$ of jet measured at the collector
as function of the dimensionless parameter $H\propto1/l_{step}$,
which increases by decreasing $l_{step}$. A trend line is drawn to guide the reader's eye and highlight the asymptotic behavior.
It is worth stressing that the radius reduction ratio of the jet is here equal
to only one order of magnitude, since in the 1-D model the bending instabilities are neglected and, therefore, the jet path is
considerably shorter than in the corresponding 3-D simulation.}

\label{Fig:rad-asy-7.2}
\end{figure}

\begin{figure}[H]
\begin{centering}
\includegraphics[scale=0.12]{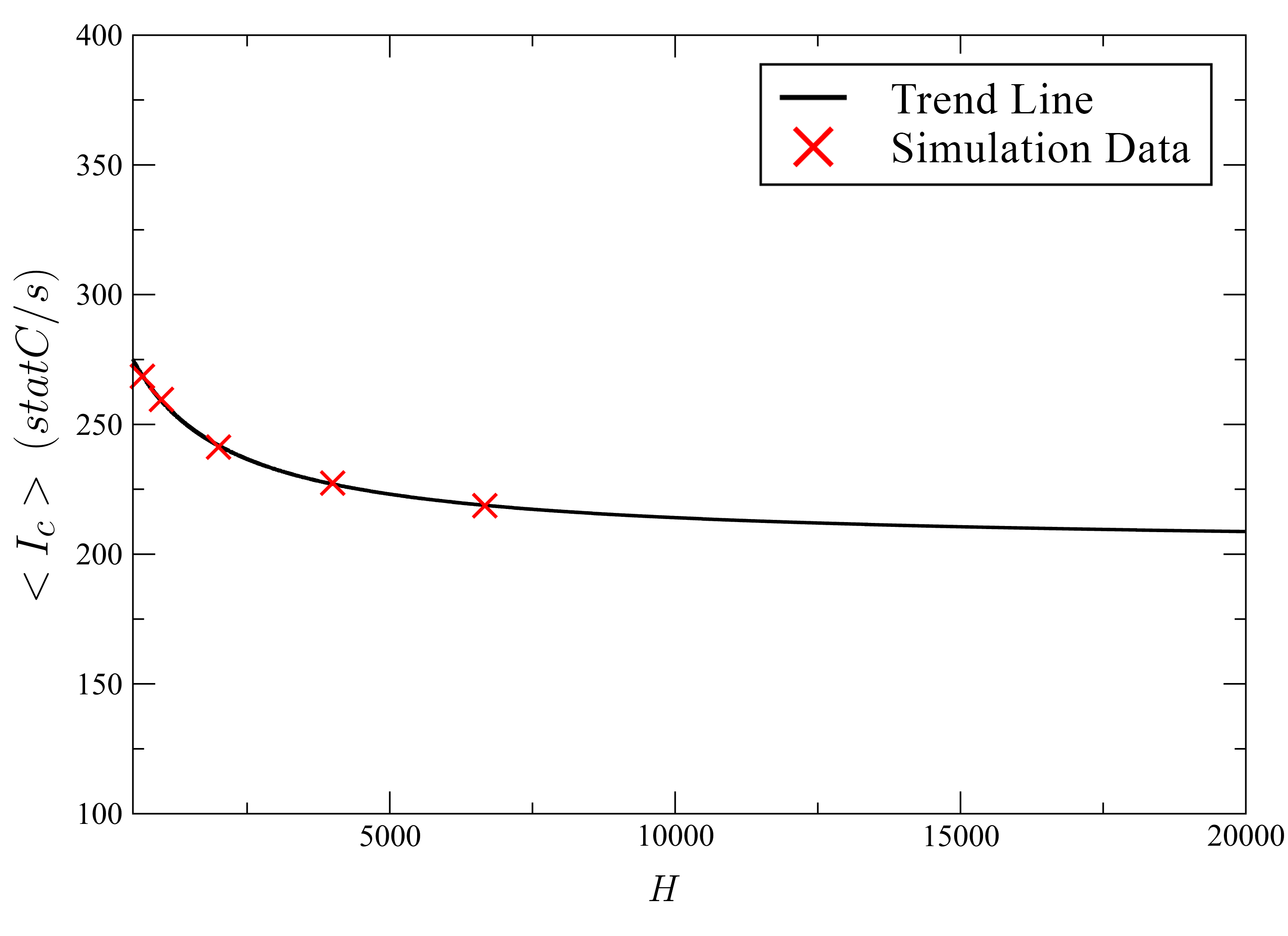}
\par\end{centering}

\protect\caption{Averaged current $<I_{c}>$ measured at the collector as function
of the dimensionless parameter $H\propto1/l_{step}$, which increases
by decreasing $l_{step}$. 
A trend line is drawn to guide the reader's eye and highlight the asymptotic behavior.
}

\label{Fig:current-asy-7.2}
\end{figure}

Here, we run the input example labeled \textbf{input 2} with different
values of length step $l_{step}$. In particular, we probe the interval
of $l_{step}$ values from $0.05$ to $0.3 \text{cm}$. All the simulations
started with the initial conditions $x=l_{step}$, $\sigma=0$, and
$\upsilon_{x}=0$, and we integrated the EOM for 2 seconds. In all
the simulations we observed an initial drift of the observables describing
the electrospinning process. This drift occurs in the early stage
of dynamics, when the jet has not reached the collector
yet. After the jet touches the collector, the observables fluctuate
around a constant mean value, providing a stationary regime. As
example, we report in Fig\ref{Fig:dragvel-7.2} the time evolution
of the dragging velocity term $\upsilon_{d}$. Note that here $\upsilon_{d}$
is not equal to zero as in the previous case, since we have activated
the injection of new beads by the directive \texttt{inserting yes}
in the input file. In Fig\ref{Fig:rad-asy-7.2} and Fig\ref{Fig:current-asy-7.2}
we report the trend of two observables, average fiber radius and current
measured at the collector in stationary regime, as function of the
dimensionless parameter $H\propto1/l_{step}$, which increases by
decreasing $l_{step}$ (see Tab\ref{tab:tabella-adimensionale} and
Eq \ref{eq:Length-scale}). In particular, we observe a slow asymptotic
behavior of the two statistical data. Further discussions on the asymptotic
behavior can be found in Refs \cite{reneker2000bending,carroll2011discretized,kowalewski2005experiments}.
It is worth stressing that the computational cost increases rapidly
by decreasing $l_{step}$, and, therefore, the user should cleverly
tune the length step $l_{step}$ in order to achieve a good compromise
between efficiency and accuracy.

\begin{figure}[H]
\begin{centering}
\includegraphics[scale=0.5]{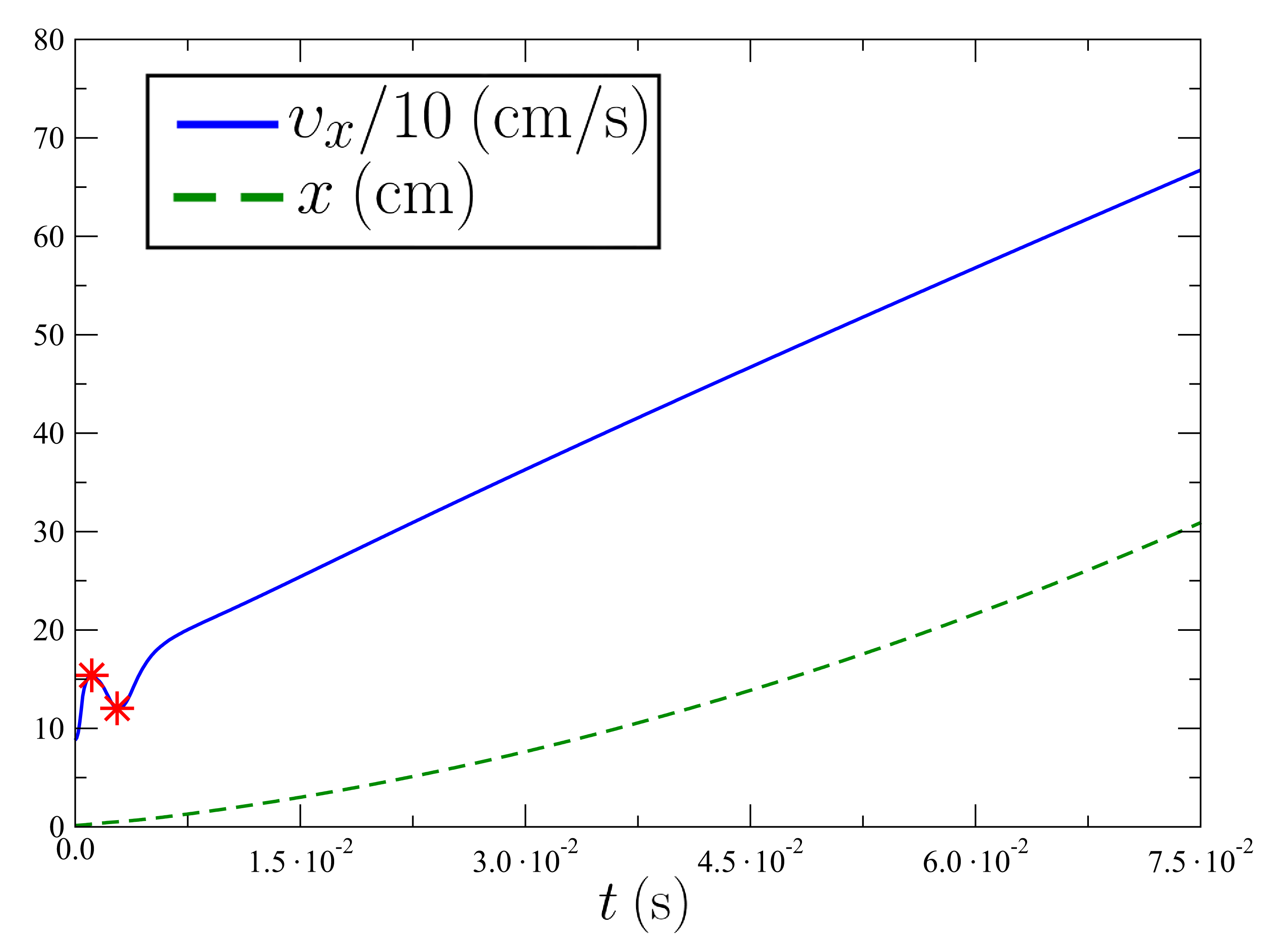}
\par\end{centering}

\protect\caption{Time evolution of the velocity $\upsilon_{x}$ rescaled by a factor
1/10 (continuous line) and the length $x$ (dashed line). Two stages
of the elongation process are observed. The first stage is biased
by the sum of viscoelastic and Coulomb forces, and shows two quasi
stationary points (denoted by red star symbols). Then, in the second
stage the velocity comes to a near linearly increasing regime under
the effect of the external electric field.}

\label{Fig:time-evol-7.2}
\end{figure}

We report in Fig\ref{Fig:time-evol-7.2} the time evolution of the
position $x$ and velocity $\upsilon_{x}$ for a generic jet segment
(jet bead) falling from the nozzle in stationary regime. Similarly
to the previous case, we identify two sequential stages in the elongation
process: the first biased by the sum of the viscoelastic and Coulomb
forces, and the second dominated by the external electric field. Here the
main difference compared to the previous case is the presence
of two quasi stationary points instead of one. This fact is due to
the larger viscoelastic force exerted by the new jet segments (beads)
inserted at the nozzle. In fact, the viscoelastic force has a braking
effect between the two quasi stationary points. Furthermore, comparing
with the previous case we observe here that the presence of a non
zero dragging velocity term $\upsilon_{d}$ anticipates the second
stage.

\subsection{Three-D simulations of a process leading to polymer nanofibers}

The electrospinning of polyvinylpyrrolidone (PVP) nanofibers is a
prototypical process, which has been largely investigated in literature.\cite{yarin2014fundamentals,pisignanoelectrospinning,persano2013industrial}
Recently, its process dynamics was experimentally probed at an ultra-high
time rate resolution by Montinaro et al.\cite{montinaro2015electrospinning}
Here, we simulate the electrospinning process of PVP solutions. Then,
the theoretical results predicted by the models are compared with
the aforementioned experimental data. In particular, we reproduce an experiment
in which a solution of PVP (molecular weight = 1300 kDa)
is prepared by a mixture of ethanol and water (17:3 v:v), at a concentration
of about 2.5 wt\%. The applied voltage is in a range around 10 kV, and the collector is placed at
distance 16 cm from the nozzle, which has radius $250\,$micron
(further details are provided in Ref. \cite{montinaro2015electrospinning}
). As rheological properties of such system we consider the zero-shear
viscosity $\mu_{0}=0.2\, \text{g}/ (\text{cm} \cdot\text{s})$ 
\cite{yuya2010morphology,buhler2005polyvinylpyrrolidone},
the elastic modulus $G=5 \cdot 10^{4} \, \text{g}/ (\text{cm} \cdot \text{s}^{2} )$ \cite{morozov2012water},
and the surface tension $\alpha=21.1\, \text{g}/ \text{s}^{2}$ \cite{yuya2010morphology}.
We use for the simulation a viscosity value $\mu$ which
is two order of magnitude larger than the zero-shear viscosity $\mu_{0}$
reported, since the strong longitudinal flows we are dealing with can lead to an increase of the extensional viscosity
from $\mu_{0}$, as already observed in literature.\cite{reneker2000bending,yarin1993free}
The mass density is equal to $0.84\,\text{g}/\text{cm}^{3}$, while
the charge density was estimated by experimental observations of the
current measured at the nozzle. For convenience, all the simulation
parameters are summarized in Tab\ref{tab:simulation-param}. 

We show here the results three independent simulations with different values of
potential $V_{0}$ between the nozzle and the collector: 6 kV, 9 kV
and 11 kV, respectively. The simulations were carried out with a time step of $10^{-7}$ s, the EOM were integrated in time for 20 million steps by
using the second order accurate Heun scheme. As example, one of the
three input files is reported as labeled \textbf{input 3}. Note that
we have activated the perturbation module by the directive
\texttt{perturb yes} in the input file, in order to model a mechanical perturbation at
the nozzle. Here, we use the frequency of the perturbation value $\omega$
proposed by Reneker et al.\cite{reneker2000bending}

\begin{figure}[H]
\begin{centering}
\includegraphics[scale=0.4]{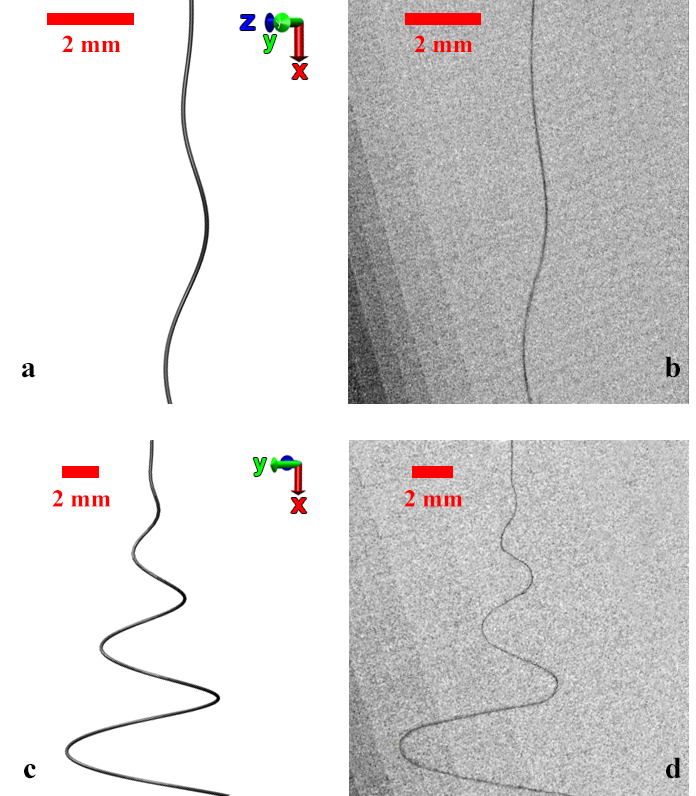}
\par\end{centering}

\protect\caption{Snapshots of the simulated jet (a,c) and of the experimental jet
(b,d) taken close the nozzle in the early stage (a,b), and in the bending
regime (c,d) of dynamics. }

\label{Fig:compare-7.3}
\end{figure}

In all the simulations, we recognize three different stages of the
electrospun jet: 1) extensione along a straight line over a few
centimeters, 2) slight perturbation from the linear path leading to bending instability; 3) fully three-dimensional
motion out from the stretching axis. Two snapshots of the simulated jet are reported on the
left of Fig\ref{Fig:compare-7.3}, which can be compared with two high-frame-rate micrographs
(on the right of the figure) collected during electrospinning
experiments \cite{montinaro2015electrospinning}. The two shown snapshots correspond to an early stage and a later regime of instabilities, respectively.
We note a general good agreement
between the simulation results and experimental measurements. Furthermore,
we monitored the velocity of the jet ejected at the nozzle, whose
value was estimated between $2.0$ and $2.4$ m/s for the case
$V_{0}=9$kV. This is also pretty close to the experimental observations
which locate the velocity around 2.6 m/s \cite{montinaro2015electrospinning}.

\begin{figure}[H]
\begin{centering}
\includegraphics[scale=0.75]{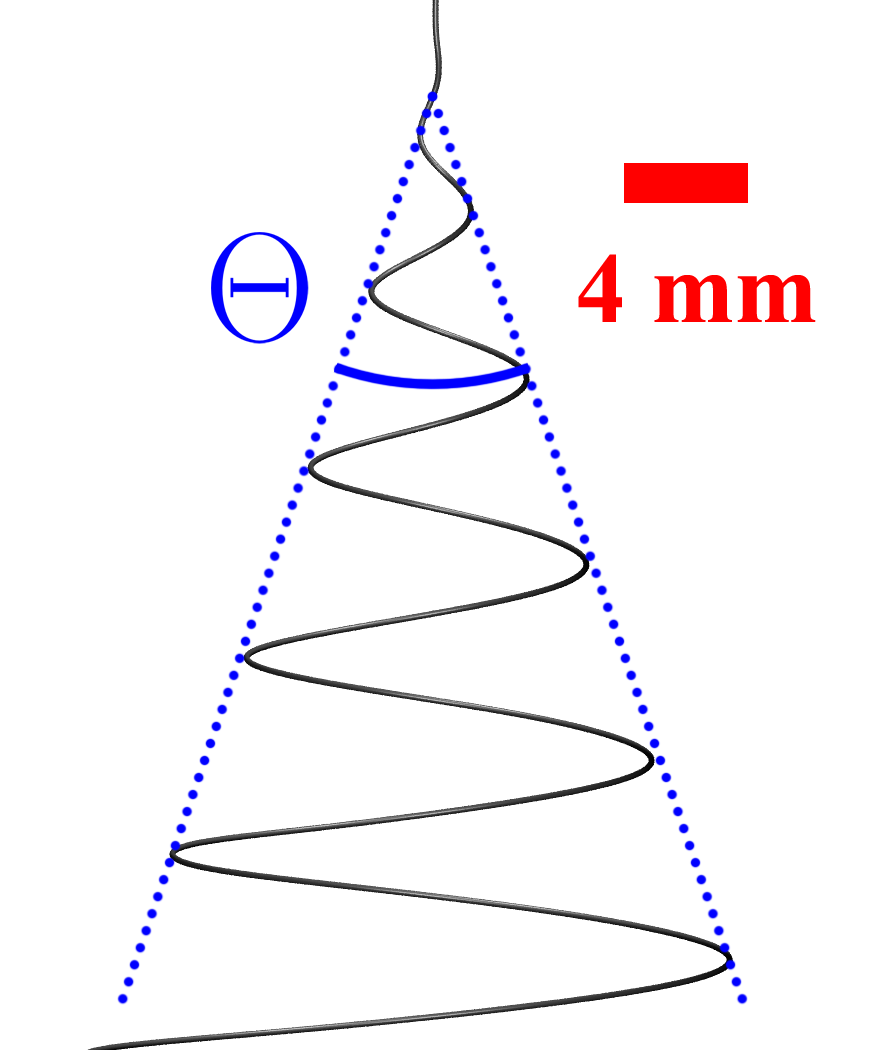}
\par\end{centering}

\protect\caption{Snapshot showing the instability broadening in stationary regime and
highlighting the resulting angular aperture $\Theta$.}

\label{fig:definizione-angolo}
\end{figure}

As mentioned above, bending instabilities play an important role in the electrospinning
process, since they increase the path traveled by the jet from the
nozzle to the collector with beneficial effect in terms of yielding a smaller polymer fiber radius. Thus, we measure
the instantaneous angular aperture ($\Theta$) of the instability
cone (see Fig\ref{fig:definizione-angolo}) in the range of potential $V_{0}$ imposed between the nozzle and
the collector (from 6 up to 11 V). The $\Theta$ values are in the range 30-36 degrees, 
which is consistent with the experimentally measured range 29-37 degrees reported in Ref \cite{montinaro2015electrospinning}.

\section{Performance\label{sec:Performance}}

The performance of the underlying numerical routines in JETSPIN is
an important aspect to take into account, affecting the choice of the value of the discretization length
step allowing the jet to be discretised without loosing accuracy.
In simulating electrospinning processes, this factor is especially
important given that upon decreasing the discretization length step the
systems size (number of beads) increases with linear dependence. Furthermore,
the code currently spends of order $N^{2}$ operations to compute
the Coulomb force ($N$ is the number of beads). The last point could
be addressed in future versions of the code by using strategies like linked
cell or tree-code algorithm. 

In this Section, we compare the CPU wall-clock time required to run
the \textbf{input 1} and \textbf{input 3} with the three different
integration schemes implemented in JETSPIN. For the \textbf{input 1}
case, we note that the increase of CPU wall-clock time is strongly
dependent on the greater complexity of the chosen integrator scheme
(see Tab\ref{Tab:cpu-time}). In particular, we note that the Heun
and Runge-Kutta schemes require around two and four times
the CPU wall-clock time cost of the Euler integrator, respectively. A similar trend
is observed also for the \textbf{input 3} case. However, the user
should also consider the different numerical accuracy provided by
the three schemes, as already discussed in Subsec \ref{sub:Numerical accuracy}. 

We probe the parallel efficiency of JETSPIN. In particular, we run
the \textbf{input 3} at different values of discretization length
step and collector-nozzle distance. Thus, we probe different system
sizes corresponding to different numbers of beads, which are used
to discretise the jet. Further, the procedure is repeated with
different number of processing cores, so that the efficiency of the
implemented parallel strategy is investigated. In order to perform
a quantitative analysis, we compute the speedup ($S_{p}$) 

\begin{equation}
S_{p}=\frac{T_{s}}{T_{p}},
\end{equation}

where $T_{s}$ stands for the CPU wall-clock time of the code in serial mode,
and $T_{p}$ the CPU wall-clock time of the code in parallel mode executed by
the number of processors denoted $n_{p}$.

We also estimate the parallel efficiency ($E_{p}$) defined as

\begin{equation}
E_{p}=\frac{S_{p}}{n_{p}}.
\end{equation}

\begin{figure}[H]
\begin{centering}
\includegraphics[scale=0.12]{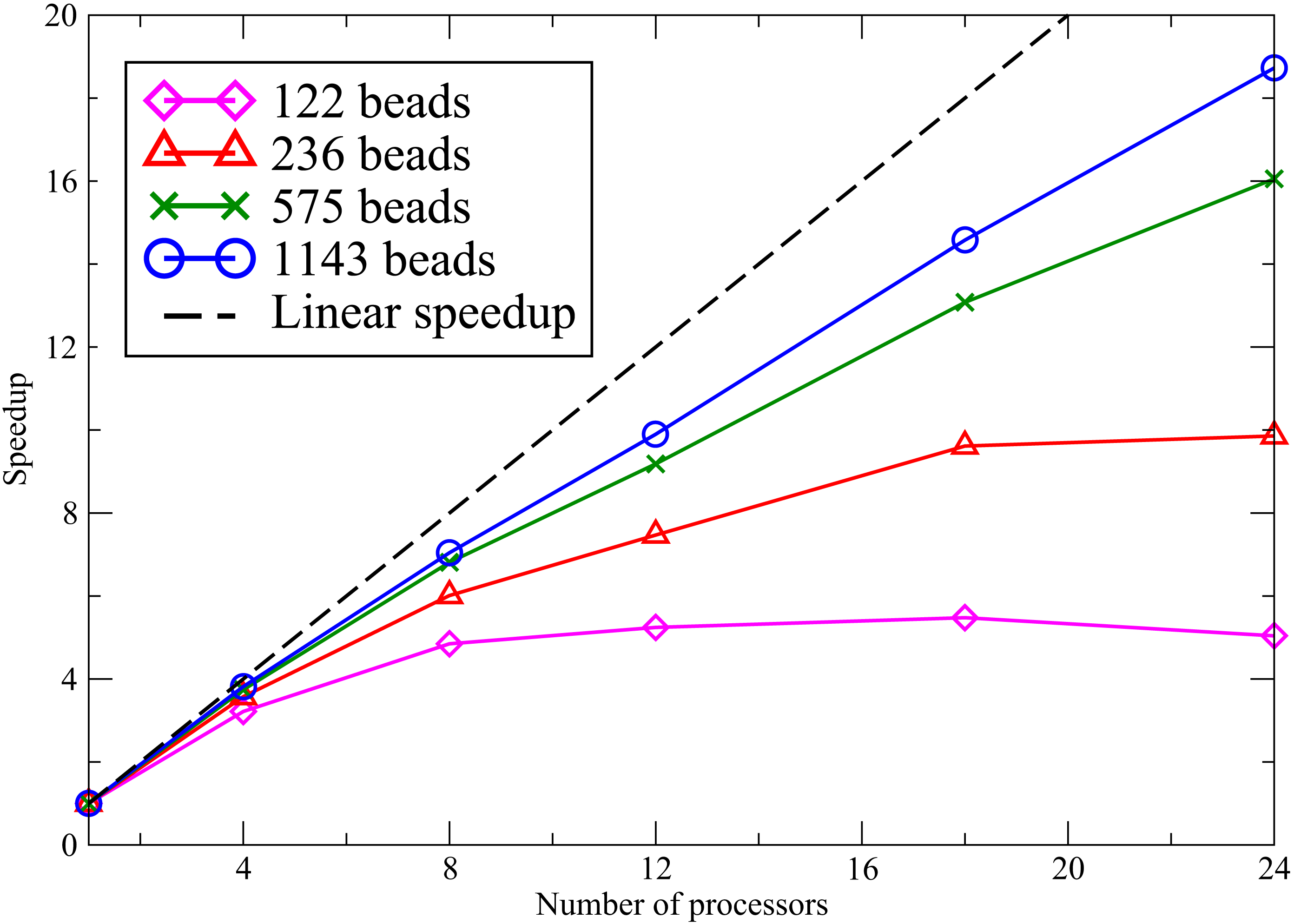}
\par\end{centering}

\protect\caption{Speedup versus number of processors for different system sizes. Ideally,
the speedup should be equal to the number of processors (linear speedup
in black dashed line). }

\label{Fig:speedup}
\end{figure}

\begin{figure}[H]
\begin{centering}
\includegraphics[scale=0.12]{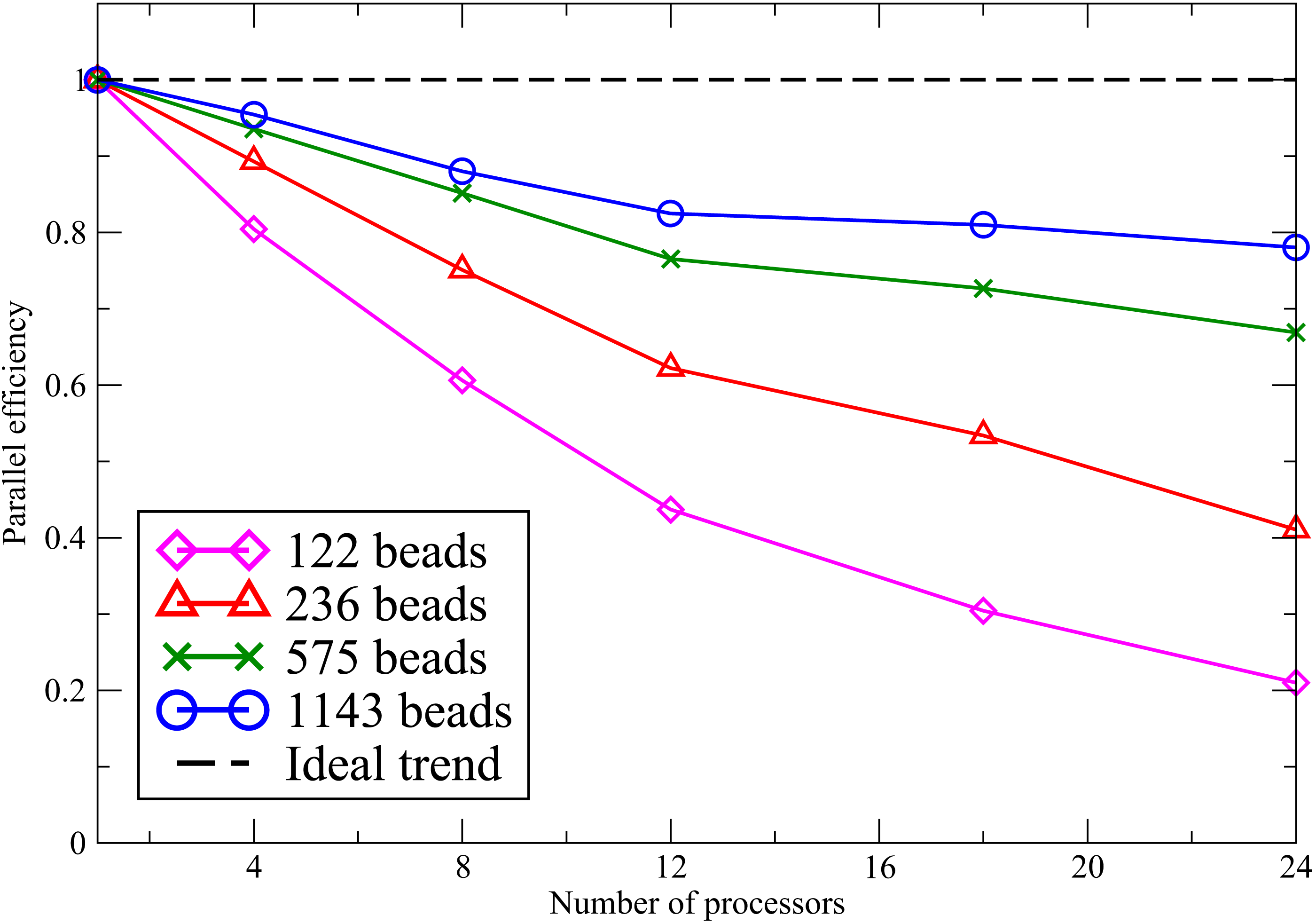}
\par\end{centering}

\protect\caption{Parallel efficiency versus number of processors for different system
sizes. Note that we observe a parallel efficiency larger than 0.8
only if the system size provides at least 50 beads per each processor.
We report the ideal parallel efficiency in black dashed line which
should be equal to one for any number of processors.}

\label{Fig:parallel-eff}
\end{figure}

The trends of these two estimators versus the number of processors
are reported in Fig\ref{Fig:speedup} and Fig\ref{Fig:parallel-eff}.
Note that only for the largest system size we observe a quasi linear
trend of $S_{p}$ (see Fig\ref{Fig:speedup}), while in all the other
cases a sub-linear behavior is evident (note that the speedup should
ideally be equal to the number of processors). This is not surprising,
since the communication latency strays actually the speedup from the
linear trend. Note that deviation from the linear trend is larger
for smaller system sizes with a lower cost-benefit ratio. This is
usually due to the communication latency which deteriorates the performance
whenever the number of beads assigned to each processor is not sufficiently
high to offset the communication overheads. 
This is  evident in Fig\ref{Fig:parallel-eff}, where
the parallel efficiency is usually larger than 0.8 (80\%) only if
the system size provides at least 50 beads per each processor. 
As a consequence, the user should always consider the number of beads
handled by each processor in order to run efficiently JETSPIN in parallel
mode.

 \section {Availability}

JETSPIN is available free under the Open Software License v. 3.0 (OSL) 
created by Lawrence Rosen\cite{osl}.
However, it is worth stressing that all commercial rights derived from this software 
are entirely owned by the authors as reported in paragraph 4 of the OSL.
A copy of the code may be obtained as zipped and tarred file at the website: http://www.nanojets.eu/.
All enquiries regarding how to obtain a copy of JETSPIN should be
addressed to the authors. 

\section {Conclusions}
We have presented JETSPIN, a new open-source software for 
the numerical simulation of electrospinning phenomena. 
JETSPIN is implemented in FORTRAN programming language and
permits to simulate a wide spectrum of  structured polymer fibers 
with diameters in the range from several micrometers down to 
tens of nanometers, which are of considerable interest for various applications. 
In this work, we have described the basic structure of the code and the
main features of the implemented model. 
In addition, we have discussed a number of examples intended to convey the reader
a flavor of the type of problems which are suitable for JETPSIN simulation. 
In particular, the simulations of prototypical polymer materials demonstrate 
the capabilities of the present code to reproduce experimental data.

JETSPIN is expevcted to provide a useful tool to investigate relationships 
among the relevant process variables,  material parameters and 
experimental settings, and to predict the dynamics and the characteristics 
of the jet, many of which can be found in resulting collected nanofibers. 
Further, it can be used to identify the role played by the concurrent physical  
mechanism participating to the electrospinning process.
In summary, JETSPIN can complement and support experiments
with the aim of enhancing the efficiency of the electrospinning
process and the quality of electrospun fiber materials.

\section*{Acknowledgments}

The software development process has received funding from the European
Research Council under the European Union's Seventh Framework Programme
(FP/2007-2013)/ERC Grant Agreement n. 306357 (\textquotedbl{}NANO-JETS\textquotedbl{}).
Miss M. Montinaro and Dr. V. Fasano are gratefully acknowledged for experimental data.

\appendix

\section*{Appendix A}

\begin{center}
\begin{tabular}{ll}
\hline 
\textbf{directive:} & \textbf{meaning:}\tabularnewline
\hline 
\hline 
\uline{collector distance} $f$ & distance of collector from the nozzle along the x axis\tabularnewline
 & \quad (note the nozzle is assumed in the origin)\tabularnewline
cutoff $f$ & length of the proximal jet sections interacting by Coulomb \tabularnewline
 & \quad force (default: equal to the collector distance)\tabularnewline
\uline{density mass} $f$ & mass density of the jet\tabularnewline
\uline{density charge} $f$ & electric volume charge density of the jet\tabularnewline
\uline{elastic modulus} $f$ & elastic modulus of the jet\tabularnewline
external potential $f$ & electric potential between the nozzle and collector\tabularnewline
\uline{final time} $f$ & set the end time of the simulation \tabularnewline
\uline{finish} & close the input file (last data record)\tabularnewline
gravity yes & activate the inclusion of the gravity force in the model\tabularnewline
\uline{initial length} $f$ & length of the jet at the initial time\tabularnewline
\uline{integrator} $i$ & set the integration scheme. The \textit{integer} can be '1' for\tabularnewline
 & \quad Euler, '2' for Heun, and '3' for Runge-Kutta scheme\tabularnewline
inserting yes & inject new beads as explained in Subsec \ref{sub:Jet-insertion} \tabularnewline
\uline{nozzle cross} $f$ & cross section radius of the jet at the nozzle\tabularnewline
nozzle stress $f$ & viscoelastic stress of the jet at the nozzle\tabularnewline
nozzle velocity $f$ & bulk fluid velocity of the jet at the nozzle\tabularnewline
perturb yes & activate the periodic perturbation at the nozzle\tabularnewline
perturb freq $f$ & frequency of the perturbation at the nozzle\tabularnewline
perturb ampl $f$ & amplitude of the perturbation at the nozzle\tabularnewline
points $i$ & number of segment in which the jet is discretised\tabularnewline
print list $s_{1}\,\ldots$ & print on terminal statistical data as indicated by symbolic\tabularnewline
 & \quad strings (see Appendix B for detailed informations)\tabularnewline
print time $f$ & print data on terminal and output file every $f$ seconds\tabularnewline
print xyz $f$ & print the trajectory in XYZ file format in centimeters\tabularnewline
 &  \quad every $f$ seconds \tabularnewline
print xyz maxnum $i$ & print the XYZ file format with $i$ beads starting from the\tabularnewline
 & \quad nearest bead to the collector (default: 100)\tabularnewline
print xyz rescale $f$ & print the XYZ file format data rescaled by $f$\tabularnewline
printstat list $s_{1}\,\ldots$ & print on output file statistical data as indicated by symbolic\tabularnewline
 & \quad strings (see Appendix B for detailed informations)\tabularnewline
removing yes & remove beads at collector\tabularnewline
resolution $f$ & discretization step length of the jet\tabularnewline
surface tension $f$ & surface tension coefficient of the jet\tabularnewline
\uline{system} $i$ & set the jet model. The \textit{integer} can be equal to \tabularnewline
 & \quad '1' for select the 1d-model or '3' for the 3d-model\tabularnewline
\uline{timestep} $f$ & set the time step for the integration scheme\tabularnewline
\uline{viscosity} $f$ & viscosity of the jet\tabularnewline
\hline 
\end{tabular}
\par\end{center}

Here, we report the list of directives available in JETSPIN. Note
$i$, $f$, and $s$ denote an integer number, a floating point number,
and a string, respectively. The underlined directives are mandatory.
The default value is zero (exceptions are stressed in parentheses).

\section*{Appendix B}

\begin{center}
\begin{tabular}{ll}
\hline 
\textbf{keys:} & \textbf{meaning:}\tabularnewline
\hline 
\hline 
t & unscaled time\tabularnewline
ts & scaled time \tabularnewline
x, y, z & unscaled coordinates of the farest bead \tabularnewline
xs, ys, zs & scaled coordinates of the farest bead\tabularnewline
st & unscaled stress of the farest bead\tabularnewline
sts & scaled stress of the farest bead\tabularnewline
vx, vy, vz & unscaled velocities of the farest bead \tabularnewline
vxs, vys, vzs & scaled velocities of the farest bead \tabularnewline
yz & unscaled normal distance from the x axis of the farest bead\tabularnewline
yzs & scaled normal distance from the x axis of the farest bead\tabularnewline
mass & last inserted mass at the nozzle \tabularnewline
q & last inserted charge at the nozzle\tabularnewline
cpu & time for every print interval\tabularnewline
cpur & remaining time to the end \tabularnewline
cpue & elapsed time \tabularnewline
n & number of beads used to discretise the jet\tabularnewline
f & index of the first bead \tabularnewline
l & index of the last bead \tabularnewline
\uline{curn} & current at the nozzle \tabularnewline
\uline{curc} & current at the collector \tabularnewline
\uline{vn} & velocity modulus of jet at the nozzle \tabularnewline
\uline{vc} & velocity modulus of jet at the collector\tabularnewline
\uline{svc} & strain velocity at the collector \tabularnewline
\uline{mfn} & mass flux at the nozzle \tabularnewline
\uline{mfc} & mass flux at the collector \tabularnewline
\uline{rc} & radius of jet at the collector \tabularnewline
\uline{rrr} & radius reduction ratio of jet \tabularnewline
\uline{lp} & length path of jet \tabularnewline
\uline{rlp} & length path of jet divided by the collector distance\tabularnewline
\hline 
\end{tabular}
\par\end{center}

In JETSPIN a series of instantaneous and statistical data are available
to be printed by selecting the appropriate key. Here, the list of
symbolic string keys is reported with their corresponding meanings.
Note that by \textit{scaled} we mean that the observable was rescaled by
the characteristic values provided in Tab\ref{tab:tabella-adimensionale}.
The underlined keys correspond to data which are averaged on the time
interval given by the directive \texttt{print time} in input file.
By \textit{farest bead} we mean the farest bead from the nozzle
(with greatest $x$ value). All the quantities are expressed
in centimeter\textendash gram\textendash second unit system, excepted
the dimensionless scaled observables.

\newpage{}

\newpage

\section*{Tables}

\begin{table}[H]
\begin{centering}
\begin{tabular}{ll}
\hline 
\textbf{target:}  & \textbf{meaning:}\tabularnewline
\hline 
\hline 
gfortran & compile in serial mode using the GFortran compiler.\tabularnewline
gfortran-mpi & compile in parallel mode using the GFortran compiler and the Open
Mpi library.\tabularnewline
cygwin & compile in serial mode using the GFortran compiler under the command-line\tabularnewline
 & interface Cygwin for Windows.\tabularnewline
cygwin-mpi & compile in parallel mode using the GFortran compiler and the Open
Mpi library\tabularnewline
 & under the command-line interface Cygwin for Windows (note a precompiled\tabularnewline
 & package of the Open Mpi library is already available on Cygwin).\tabularnewline
intel & compile in serial mode using the Intel compiler.\tabularnewline
intel-mpi & compile in parallel mode using the Intel compiler and the Intel Mpi
library.\tabularnewline
intel-openmpi & compile in parallel mode using the Intel compiler and the Open Mpi
library.\tabularnewline
help & return the list of possible target choices\tabularnewline
\hline 
\end{tabular}
\par\end{centering}

\protect\caption{List of targets for several common workstations and parallel computers,
which can be used by the command "\texttt{make target}".}

\label{Tab:targets}
\end{table}

\begin{table}[H]
\begin{centering}
\begin{tabular}{cc}
\hline 
\multicolumn{2}{c}{Characteristic Scales}\tabularnewline
\hline 
\hline 
$L_{0}=l_{step}\sqrt{\frac{\pi a_{0}^{2}\rho_{V}^{2}}{G}}$ & $t_{0}=\frac{\mu}{G}$\tabularnewline
\multicolumn{2}{c}{$\sigma_{0}=G$}\tabularnewline
\hline 
\multicolumn{2}{c}{Dimensionless Derived Variables}\tabularnewline
\hline 
\hline 
$\bar{l}_{i}={\displaystyle \frac{l_{i}}{L_{0}}}$ & $\bar{R}_{ij}={\displaystyle \frac{R_{ij}}{L_{0}}}$\tabularnewline
\multicolumn{2}{c}{$\bar{k}=kL_{0}$}\tabularnewline
\hline 
\multicolumn{2}{c}{Dimensionless Groups}\tabularnewline
\hline 
\hline 
$V_{i}={\displaystyle \frac{q_{i}V_{0}\mu^{2}}{m_{i}h\, L_{0}\, G^{2}}}$ & $Q_{ij}={\displaystyle \frac{q_{i}q_{j}\mu^{2}}{L_{0}^{3}m_{i}G^{2}}}$\tabularnewline
$F_{ve,i}={\displaystyle \frac{\pi a_{0}^{2}\mu^{2}}{m_{i}L_{0}G}}$  & $A_{i}={\displaystyle \frac{\alpha\pi a_{0}^{2}\mu^{2}}{m_{i}L_{0}^{2}G^{2}}}$\tabularnewline
$F_{g}={\displaystyle \frac{g\mu^{2}}{L_{0}G^{2}}}$ & $K_{s}=\omega{\displaystyle \frac{\mu}{G}}$\tabularnewline
$H={\displaystyle \frac{h}{L_{0}}}$ & $L_{step}={\displaystyle \frac{l_{step}}{L_{0}}}$\tabularnewline
\hline 
\end{tabular}
\par\end{centering}

\protect\caption{Definitions of the characteristic scales, dimensionless derived variables,
and groups employed in the text.}

\label{tab:tabella-adimensionale}
\end{table}

\begin{table}[H]
\begin{centering}
\begin{tabular}{cccccccccc}
\hline 
$\rho$ & $\rho_{q}$ & $a_{0}$ & $\upsilon_{s}$ & $\alpha$ & $\mu$ & $G$ & $V_{0}$ & $\omega$ & $A$\tabularnewline
($\text{kg}/\text{m}^{3}$) & ($\text{C}/\text{L}$) & ($\text{cm}$) & ($\text{cm/s}$) & (mN/m) & (Pa$\cdot$s) & (Pa) & (kV) & ($\text{s}^{-1}$) & ($\text{cm}$)\tabularnewline
\hline 
\hline 
840 & $2.8\cdot10^{-7}$ & $5\cdot10^{-3}$ & 0.28 & 21.1 & 2.0 & 50000 & 9.0 & $10^{4}$ & $10^{-3}$\tabularnewline
\hline 
\end{tabular}
\par\end{centering}

\protect\caption{Simulation parameters for the simulation of PVP nanofibers. The headings
used are as follows: $\rho$: density, $\rho_{q}$: charge density,
$a_{0}$: fiber radius at the nozzle, $\upsilon_{s}$: bulk fluid
velocity in the syringe needle, $\alpha$: surface tension, $\mu$:
viscosity, $G$ : elastic modulus, $V_{0}$ : applied voltage bias, $\omega$:
frequency of perturbation, $A$ : amplitude of perturbation. The bulk fluid velocity $\upsilon_{s}$ was estimated considering
that the solution was pumped at constant flow rate of 2 mL/h in a
needle of radius $250\,$micron.}

\label{tab:simulation-param}
\end{table}

\begin{table}[H]
\begin{centering}
\begin{tabular}{ccccccc}
\hline 
Input file & Integrator scheme & \# of CPUs & Parallel & CPU time (s)* & \# of beads & CPU time (s)*\tabularnewline
 &  &  & efficiency &  &  & per bead and step\tabularnewline
\hline 
\hline 
\textbf{Input 1} & Euler  & 1 & - & 4.76 & 1 & $4.76\cdot10^{-7}$\tabularnewline
\textbf{Input 1} & Heun & 1 & - & 8.64 & 1 & $8.64\cdot10^{-7}$\tabularnewline
\textbf{Input 1} & Runge-Kutta & 1 & - & 16.72 & 1 & $16.72\cdot10^{-7}$\tabularnewline
\textbf{Input 3} & Euler & 4 & 0.8 & 11307.47 & 150 & $1.51\cdot10^{-6}$\tabularnewline
\textbf{Input 3} & Heun & 4 & 0.8 & 22368.07 & 150 & $2.98\cdot10^{-6}$\tabularnewline
\textbf{Input 3} & Runge-Kutta & 4 & 0.8 & 44708.62 & 150 & $5.96\cdot10^{-6}$\tabularnewline
\hline 
\end{tabular}
\par\end{centering}

\protect\caption{We report the CPU wall-clock time in seconds which is needed
to run the \textbf{input 1} and \textbf{input 3} example files. For
each of the two input files we test the three different integration
schemes implemented in JETSPIN: the first-order accurate Euler scheme,
the second-order accurate Heun scheme, and the fourth-order accurate
Runge-Kutta scheme. *The benchmark of the \textbf{input 1} was carried
out on an Intel Core I5 480M (3M Cache, 2.66 GHz) in serial mode,
while the \textbf{input 3} benchmark was executed in parallel mode
(4 CPUs) on a node of 2x12 core processors made of 2.4 GHz Intel Ivy
Bridge cores. We report also the number of beads used to discretise
the jet, and the CPU time which is needed to integrate one bead
for one time step.}

\label{Tab:cpu-time}
\end{table}

\begin{table}[H]
\begin{centering}
\begin{tabular}{l}
\hline 
\textbf{Input 1.} Example JETSPIN input file for electrospinning simulation.\tabularnewline
\hline 
\hline 
\texttt{system 1}\tabularnewline
\texttt{integrator 2}\tabularnewline
\texttt{timestep 1.d-8}\tabularnewline
\texttt{final time 1.d-1}\tabularnewline
\texttt{print time 1.d-3}\tabularnewline
\texttt{print list ts xs sts vxs cpu cpur cpu}\tabularnewline
\texttt{printstat list ts xs sts vxs cpu cpur cpu}\tabularnewline
\texttt{inserting no}\tabularnewline
\texttt{removing no}\tabularnewline
\texttt{points 1}\tabularnewline
\texttt{initial length 3.19d-1}\tabularnewline
\texttt{nozzle cross 1.5d-2}\tabularnewline
\texttt{density mass 8.18912288d-1}\tabularnewline
\texttt{density charge 3761.26389d0}\tabularnewline
\texttt{viscosity 100.d0}\tabularnewline
\texttt{elastic modulus   10000.d}\tabularnewline
\texttt{collector distance 200.d0}\tabularnewline
\texttt{external potential 277.8141d0}\tabularnewline
\texttt{finish}\tabularnewline
\hline 
\end{tabular}
\par\end{centering}

\label{Tab:input-file-7.1}
\end{table}

\begin{table}[H]
\begin{centering}
\begin{tabular}{l}
\hline 
\textbf{Input 2.} Example JETSPIN input file for electrospinning simulation.\tabularnewline
\hline 
\hline 
\texttt{system 1}\tabularnewline
\texttt{integrator 3}\tabularnewline
\texttt{timestep 1.d-7}\tabularnewline
\texttt{final time 2.d0}\tabularnewline
\texttt{print time 1.d-2}\tabularnewline
\texttt{print list t x cpu cpur n rc curc}\tabularnewline
\texttt{printstat list t x st vx cpu cpue cpur n rc curc vc mfc}\tabularnewline
\texttt{inserting yes}\tabularnewline
\texttt{removing yes}\tabularnewline
\texttt{points 1}\tabularnewline
\texttt{initial length 0.10d0}\tabularnewline
\texttt{nozzle cross 1.5d-2}\tabularnewline
\texttt{density mass 8.1991d-1}\tabularnewline
\texttt{density charge 37607.35d0}\tabularnewline
\texttt{viscosity   100.d0}\tabularnewline
\texttt{elastic modulus   10000.d}\tabularnewline
\texttt{collector distance 200.d0}\tabularnewline
\texttt{external potential 277.8141d0}\tabularnewline
\texttt{finish}\tabularnewline
\hline 
\end{tabular}
\par\end{centering}

\label{Tab:input-file-7.2}
\end{table}

\begin{table}[H]
\begin{centering}
\begin{tabular}{l}
\hline 
\textbf{Input 3.} Example JETSPIN input file for electrospinning simulation.\tabularnewline
\hline 
\hline 
\texttt{system 3}\tabularnewline
\texttt{integrator 2}\tabularnewline
\tabularnewline
\texttt{timestep 1.d-8}\tabularnewline
\texttt{final time 0.5d0}\tabularnewline
\tabularnewline
\texttt{print time 1.d-2}\tabularnewline
\texttt{print list t x vn vc yz n curn curc}\tabularnewline
\texttt{printstat list t x vn vc yz n curn curc}\tabularnewline
\texttt{print xyz 1.d-4}\tabularnewline
\texttt{print xyz maxnumber 400}\tabularnewline
\tabularnewline
\texttt{inserting yes}\tabularnewline
\texttt{removing yes}\tabularnewline
\texttt{gravity yes}\tabularnewline
\tabularnewline
\texttt{points 1}\tabularnewline
\texttt{initial length 0.02.d0}\tabularnewline
\texttt{nozzle cross 5.d-3}\tabularnewline
\texttt{density mass 0.84d0}\tabularnewline
\texttt{density charge 44000.d0}\tabularnewline
\texttt{viscosity   20.d0}\tabularnewline
\texttt{elastic modulus   50000.d0}\tabularnewline
\texttt{collector distance 16.d0}\tabularnewline
\texttt{external potential 30.02d0}\tabularnewline
\texttt{surface tension 21.13d0}\tabularnewline
\texttt{perturbation yes}\tabularnewline
\texttt{perturbation frequency  1.d+4}\tabularnewline
\texttt{perturbation amplitude  1.d-3}\tabularnewline
\tabularnewline
\texttt{finish}\tabularnewline
\hline 
\end{tabular}
\par\end{centering}

\label{Tab:input-file-7.3}
\end{table}

\newpage{}

\section*{Figures}

\end{document}